\documentclass[aps,prr, onecolumn,superscriptaddress]{revtex4-2}
\usepackage[T1]{fontenc}
\usepackage{amsmath,amssymb,amsthm,mathtools}
\usepackage{graphicx}
\usepackage{enumitem}

\usepackage[colorlinks=true,citecolor=blue,linkcolor=blue]{hyperref}

\newcommand{\C}{\mathbb{C}}

\newcommand{\del}{\partial}
\newcommand{\de}{\delta}
\newcommand{\eps}{\varepsilon}
\newcommand{\beq}{\begin{equation}}
\newcommand{\eeq}{\end{equation}}
\DeclareMathOperator{\Tr}{Tr}
\DeclareMathOperator{\sign}{sign}

\newcommand{\arXiv}[1]{\href{http://www.arXiv.org/abs/#1}{arXiv:#1}}
\renewcommand{\doi}[2]{\href{http://dx.doi.org/#2}{#1}}

\begin{document}
\title{Random Matrix Theory of Sparse Neuronal Networks with Heterogeneous Timescales}
\author{Thiparat Chotibut}
\email{thiparatc@gmail.com}
\affiliation{Chula Intelligent and Complex Systems, Department of Physics, Faculty of Science, Chulalongkorn University, Bangkok, Thailand}
\author{Oleg Evnin}
\email{oleg.evnin@gmail.com}
\affiliation{High Energy Physics Research Unit, Faculty of Science, Chulalongkorn University, Bangkok, Thailand}
\affiliation{Theoretische Natuurkunde, Vrije Universiteit Brussel (VUB) and International Solvay Institutes, Brussels, Belgium}
\affiliation{Mark Kac Center for Complex Systems Research, Jagiellonian University, Krak\'ow, Poland}\
\author{Weerawit Horinouchi}
\email{wee.hori@gmail.com}
\affiliation{Department of Mathematics, King's College London, London, UK}
\affiliation{Chula Intelligent and Complex Systems, Department of Physics, Faculty of Science, Chulalongkorn University, Bangkok, Thailand}
\begin{abstract}

Training recurrent neuronal networks consisting of excitatory (E) and inhibitory (I) units with additive noise for working memory computation slows and diversifies inhibitory timescales, leading to improved task performance that is attributed to emergent marginally stable equilibria [PNAS 122 (2025) e2316745122]. Yet the link between trained network characteristics and their roles in shaping desirable dynamical landscapes remains unexplored. Here, we investigate the Jacobian matrices describing the dynamics near these equilibria and show that they are \emph{sparse, non-Hermitian rectangular-block} matrices modified by \emph{heterogeneous} synaptic decay timescales and activation-function gains. We specify a random matrix ensemble that faithfully captures the spectra of trained Jacobian matrices, arising from the \emph{inhibitory core---excitatory periphery} network motif (pruned E weights, broadly distributed I weights) observed post-training. An analytic theory of this ensemble is developed using statistical field theory methods: a \emph{Hermitized resolvent} representation of the spectral density is processed with a \emph{supersymmetry-based} treatment in the style of Fyodorov and Mirlin. In this manner, an analytic description of the spectral edge is obtained, relating statistical parameters of the Jacobians (sparsity, weight variances, E/I ratio, and the distributions of timescales and gains) to \emph{near-critical} features of the equilibria essential for robust working memory computation. 

\end{abstract}

\maketitle


\section{Introduction}
Complex systems, from neural to ecological networks, often exhibit a rich heterogeneity of timescales, with some components evolving orders of magnitude slower than others \cite{Murray2014,Bernacchia2011,Hastings2016,Garcia2019,Wander2024}. In ecology \cite{May1972,Allesina2012,Grilli2016}, heterogeneity can arise from unequal interaction strengths, shaping timescales of competition and cooperation, and influencing global ecosystem stability \cite{Jackson2023}. In the brain, cortical neurons express a broad range of intrinsic timescales, spanning milliseconds to seconds \cite{Runyan2017}, and separation of these timescales has been linked to dynamical stabilization in recurrent networks \cite{Can2025}. 
More generally, achieving marginal stability in high-dimensional complex systems is especially relevant for information processing, as operating near a stability boundary provides access to slow relaxation timescales (i.e., extended memory lifetimes for past inputs), which in turn supports robust temporal memory \cite{Ganguli2008,Toyoizumi2011,Martinez2021}.

It has long been hypothesized that operating near the \emph{edge-of-stability} endows neural networks with strong computational capabilities \cite{Bertschinger2004, Ganguli2008, Barzon2025,stabilityrecent1,stabilityrecent2}. Near this edge, dynamical behaviors are sufficiently stable to retain information yet rich enough to respond to external drives.  Classic random neural network theory identifies a stability-chaos transition as synaptic gain crosses a threshold \cite{Sompolinsky1988}; in classical and quantum reservoir computing, memory capacity peaks in some memory tasks when the reservoir is about to cross the instability threshold \cite{Jaeger2001,Maass2002, Martinez2021, Cheamsawat2025}. However, in uniformly random networks, the critical regime is \emph{fragile} and requires parameter fine-tuning. Certain network architectures can mitigate this fragile fine-tuning problem: hierarchical modular networks, for example, couple fast local dynamics to slower intermodule dynamics and can self-organize near criticality \cite{Kusmierz2025}. Likewise, frozen stabilization---a principle where a portion of a network's degrees of freedom are spontaneously frozen---produces robust marginal modes and a broad relaxation spectrum without the need for precise tuning \cite{Can2025}. Together, these findings demonstrate that specific network motifs and timescale heterogeneity can robustly realize near-marginal dynamics, with favorably slow or soft modes for temporal information processing \cite{Ganguli2008, Russo2024, Murugan2025}.

While these theories support edge-of-stability computation, a key question is whether such motifs can arise naturally in biologically plausible neural networks trained on realistic tasks. Recent work shows that excitatory-inhibitory (E/I) recurrent neuronal networks trained on working memory tasks, in the presence of additive noise ubiquitous in the cortex, self-organize to \emph{near-marginal} task-evoked discrete attractors \cite{Rungratsameetaweemana2025}. Inhibitory dynamics acquire significantly longer and more heterogeneous  timescales than excitatory ones, and this timescale heterogeneity shifts the local Jacobian spectra at an operating attractor toward the stability boundary, accompanied with improved memory performance. These findings highlight a functional role for slow, distributed inhibitory timescales in placing networks near criticality. However, the influence of  connectivity motifs (e.g. sparsity, Dale’s law with E/I ratio of $\sim$80/20 in mammalian cortex \cite{Wei2022, Pena2020, Hendry1987, Markram2004, Isaacson2011}, weight distribution), together with a reservoir of synaptic timescales, on the Jacobian spectrum remains unexplored. This motivates an analytic treatment of the local Jacobian to determine how these trained network characteristics place the spectral edge near the imaginary axis and enable robust working memory computation.

To ground the presentation for more mathematically-minded readers, let us rephrase the above qualitative descriptions as a concrete dynamical system. Configurations of a neuronal network of $N$ neurons are described by real-valued activation variables $x_i$ with $i=1,2,\cdots,N$. These variables undergo evolution according to a first-order stochastic differential equation, where the time derivative $\dot x_i$ is expressed through the influence of the other neurons on the excitation of the $i$th neuron. These influences are encoded in the weight matrix $W_{ij}$ that captures the effect of neuron $j$ upon neuron $i$. The source neuron $j$ is \emph{excitatory} (E) if it raises the activation of its target neurons ($W_{ij}> 0$) and \emph{inhibitory} (I) otherwise ($W_{ij}< 0$). Furthermore, the synaptic output generated by a neuron in the activation state $x_j$ is given by a nonlinear activation function $\sigma(x_j)$, which rapidly approaches $0$ as $x_j\to-\infty$ and saturates at a constant value as $x_j\to +\infty$. In the absence of external inputs, the activation of each neuron decays to $0$, encoded in the $-x_i$ contribution to $\dot x_i$. This leads, schematically, to the following equation of motion:
\begin{equation}
\label{schematic}
\tau_i \dot x_i(t) = - x_i(t) + \sum_{j=1}^{N} W_{ij}\sigma\big(x_j(t)\big) + \mathrm{inputs} + \mathrm{noise},
\end{equation}
where the precise form of the input signal and the noise will be given in our technical presentation in section~\ref{sec:wm_attractors}. Note the presence of the individual neuron-specific timescales $\tau_i$.

While the network is being trained (also in the presence of noise, as in \cite{Rungratsameetaweemana2025}),
the values of $W_{ij}$ and $\tau_i$ (as well as the way the input signal is distributed among the neurons and the way the final state of the network after it has responded to the stimulus is decoded into the output)
are all changed using gradient-descent-type algorithms until optimal performance has been attained.
Thereafter, these parameters remain frozen, and the operation of the network is in the form of running the dynamical evolution of (\ref{schematic}) for a given input signal.

For fixed values of parameters, (\ref{schematic}) is a large system of complicated nonlinear differential equations (due to the presence of the nonlinear activation function $\sigma$) and one expects that,
in the absence of input and noise, it possesses a complex landscape of equilibria, or `fixed points,' where $\dot x_i=0$.
(A recent discussion of such equilibria can be found in \cite{equilibria}.) As external stimuli die out,
the network settles down to one of the equilibria, and successful operation of the network depends
on its ability to reach distinct, identifiable equilibria in response to distinct stimuli. In some situations,
even the intermediate network dynamics can be visualized as successive jumps between different equilibria within the landscape. Under such circumstances, the stability properties of these equilibrium points
become of crucial importance, in particular, the presence of weakly damped, near-marginal stable modes (which place the system at the \emph{edge-of-stability}), as these slowest modes play a defining role in the long-time dynamics. Such stability questions are addressed by linearizing the dynamical system (\ref{schematic}) near each particular equilibrium. This results in the Jacobian matrix $J$ that can be recast in the form
\begin{equation}\label{jacodef}
J = \mathcal{T}^{-1}\left(-I + W  \mathcal{H}^*\right),
\end{equation}
where $\mathcal{T}$ is the diagonal matrix of heterogeneous synaptic timescales $\tau_i$ and $\mathcal{H}^*$ is a diagonal `gain' matrix at a given fixed point, expressed through the derivative of $\sigma$. The weight matrix $W$ respects the E/I block structure with sign-constrained, relatively sparse, asymmetric connectivity. The structure of the Jacobian will be described more precisely in Sec.~\ref{sec: setup_RMT}, Eqs.~(\ref{eq:Jacobian_operating_point}-\ref{eq:J_component}).

In this work, we develop a sparse non-Hermitian random-matrix theory (RMT) to explain the properties
of Jacobian matrices of the form (\ref{jacodef}) naturally arising from equilibria of trained networks originating from the setup of \cite{Rungratsameetaweemana2025}. Our goal is to connect the statistical structure of the trained weight matrices and heterogeneous neuronal timescales to the Jacobian eigenvalue spectra, which determine the stability of a typical operating attractor. Empirically, the trained weight matrices are sparse and structured: by Dale’s law they partition into excitatory and inhibitory blocks, and after training most outbound excitation is switched off while inhibition dominates and is relatively sparse, forming in this manner an \emph{inhibitory core---excitatory periphery} motif. 

Even under the assumption of independently distributed entries of $\mathcal{T}$, $W$ and $\mathcal{H}^*$
the random matrix ensemble arising from (\ref{jacodef}) lies beyond classic random matrix settings. The combination of asymmetry, sparsity, two-population (E/I) structure, and element-wise (timescale and gain) heterogeneity reaches far beyond the assumptions underlying the classic Ginibre or Girko-type results \cite{Ginibre1965,Girko1985}. Prior studies address components of this problem, e.g., random E/I matrices \cite{Rajan2006}, sparse oriented graphs \cite{Metz2019}, but require further optimization and refinement for the problem we are facing. We will follow a strategy that borrows methods from quantum and statistical field theory and goes back to the works \cite{Bray1988,Rodgers1988density} relying on the replica method. Here, we shall rely on a supersymmetry-based formalism for disorder averaging in the spirit of Fyodorov-Mirlin \cite{Fyodorov1991, Mirlin_1991}, see   \cite{Akarapipattana2023Random,Akarapipattana2025Hammerstein,Akarapipattana2025Statistical,Safonova2025spectral, Safonova2025density} for recent applications
to Hermitian random matrices. For non-Hermitian random matrices, these techniques must be applied to the Hermitized resolvent representation \cite{Metz2019} of the spectral density, as we shall explain below. After bringing the spectral density of $J$ under analytic control, our focus will be on extracting an explicit \emph{spectral-edge} condition. This edge determines how close the bulk eigenvalues of $J$ approach the stability boundary $\Re \lambda=0$, i.e., how near-marginal the trained equilibria typically are. Our theory describes the dependence of  this edge on the key parameters—the E/I ratio, network sparsity $k$, variance structure of $W$, and the inhibitory vs.\ excitatory distributions of timescales—thereby quantitatively linking emergent network statistics to features of near-marginal discrete attractors. These considerations will account for the observed bar-and-blob spectral geometry previewed in Fig.~\ref{fig:motif_spectra} (bottom right).

\paragraph*{Outline.}
The paper is organized as follows. Section~\ref{sec:wm_attractors} reviews the working memory computation and the trained E/I networks of \cite{Rungratsameetaweemana2025}, and extracts the empirical statistics to formally define the relevant random matrix ensemble. Section~\ref{sec:intro_susy} provides a pedagogical introduction to the analysis of sparse non-Hermitian RMT using the superintegral approach. Section~\ref{sec:jacobian-SUSY} applies this approach to the heterogeneous Jacobian ensemble of interest and derives the self-consistent equations for the spectral density. Specifically, we obtain an analytic description of the spectral edge in Sec.~\ref{sec:rho_I} and quantify how the interplay of sparsity, weight variances, and timescale distributions determines the distance to the stability boundary in Secs.~\ref{sec:homo-tau} and~\ref{sec:hetero-tau}. Section~\ref{sec:outlook} concludes with a discussion on the functional implications of these spectral features and outlines open problems essential for a comprehensive understanding of working memory models that use the discrete attractor hopping mechanism. Additionally, Appendix~\ref{app:RAM-like_model} highlights the utility of our framework by recovering the classic Rajan-Abbott spectral distribution \cite{Rajan2006} in the limit of dense connectivity and homogeneous timescales.

\section{From neuronal networks trained for working memory tasks\\ to a sparse non-Hermitian random matrix ensemble}
\label{sec:wm_attractors}

\begin{figure}[t]
  \centering
  \includegraphics[width=.7\linewidth]{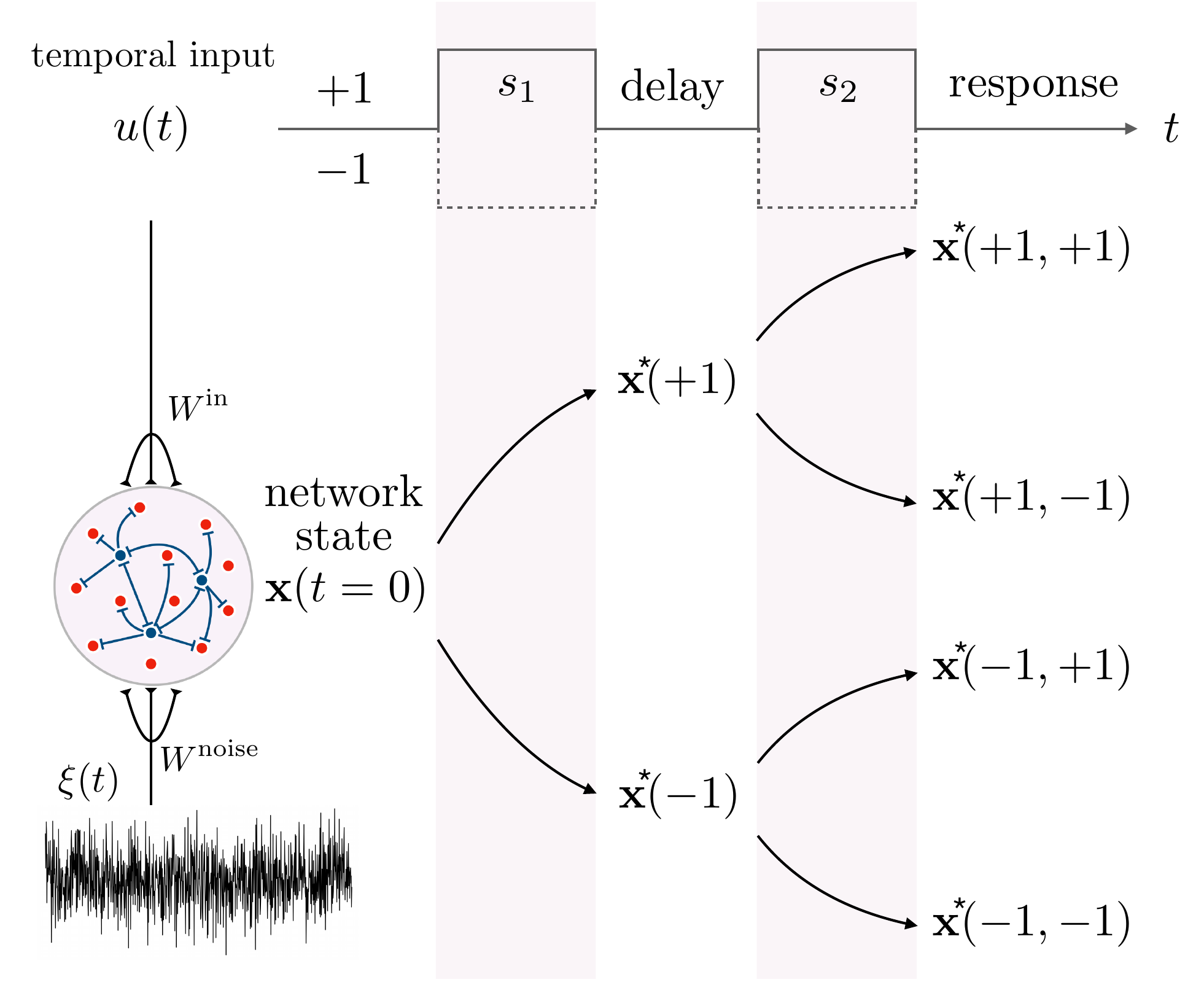}
\caption{\textbf{Input-driven transitions between discrete attractors as a paradigm of working memory computation (a delayed match-to-sample task).}
A recurrent E/I rate network governed by Eq.~\eqref{eq:dyn_model} receives binary cues $s_1, s_2 \in \{\pm1\}$ through $W^{\mathrm{in}}$ and independent noise via $W^{\mathrm{noise}}$ (bottom left). The trial timeline (top) consists of cue $s_1$, a delay, cue $s_2$, and a response window. In the network state space (bottom), $s_1$ selects one of two delay-period operating fixed points $\mathbf{x}^*(\pm1)$; the second cue then drives a controlled hop to one of four discrete attractors $\mathbf{x}^*(s_1,s_2)$. A linear readout (not shown) reports the XNOR decision (match vs. non-match) from the firing pattern of the final attractor, $\sigma(\mathbf{x}^*(s_1,s_2))$. Reliable working-memory performance in this dynamical landscape requires these operating points to be \emph{near-marginal}, stable enough during the delay to encode input history yet easily driven by the next input to hop to the subsequent operating point's basin of attraction. The emergent network motifs underlying near-marginal dynamical landscapes and their typical Jacobian spectra evaluated around operating points are shown in Fig.~\ref{fig:motif_spectra}.}
\label{fig:wm_sketch}
\end{figure}

We begin with a brief overview of the dynamical landscape that emerges in our trained recurrent neuronal networks (see Fig.~\ref{fig:wm_sketch}), then set up the sparse non-Hermitian random matrix problem relevant for analyzing the stability of discrete attractors in this dynamical landscape. The subsequent sections will describe how one should treat this random matrix problem.

\subsection{Transitioning between discrete attractors as a dynamical landscape for working memory computation}\label{sec: working memory}
One mechanism by which a recurrent network can store a temporal binary sequence is via \emph{hops between discrete attractors, each representing the past input sequence} \cite{Amit1989,Vyas2020, Duncker2021, Rungratsameetaweemana2025}. Each incoming bit drives the state out of its current basin and into a new one, thereby encoding the input into the network’s state. This input-driven discrete attractor switching has been observed in trained RNNs performing sequential tasks with interleaved delays \cite{Sussillo2013, Barak2013, Golub2018, mahes_NEURIPS2019, Rungratsameetaweemana2025}, and even in non-neural systems engineered for temporal sequence recognition; for instance, mechanical spring networks trained to traverse through a sequence of stable configurations \cite{Stern2020} and synthetic biochemical circuits that recognize temporal chemical signals as long-lived molecular states \cite{OBrien2019, Floyd2025}. The shared principle is that memory state is determined by which attractor the system currently occupies, so the attractor state encodes input history (schematized in Fig.~\ref{fig:wm_sketch}). Other dynamical mechanisms can also support temporal information processing in RNNs, e.g., continuous or line attractors \cite{Khona2022,maheswaranathan2019, Barak2013, Duncker2021} or temporal integration near the edge of chaos \cite{Bertschinger2004, Toyoizumi2011}, but here we focus on the discrete attractor mechanism, as this is realized by the model we study.

To ground this picture, consider a simple two-step working memory task used in \cite{Rungratsameetaweemana2025}: a delayed match-to-sample (DMS) task requiring a decision on whether two sequential cues $s_1, s_2 \in \{-1,+1\}$, interleaved by a delay period, are the same (timeline in Fig.~\ref{fig:wm_sketch}, top). This is logically an XNOR operation across time. The network must remember $s_1$ over the delay period and then report whether the incoming $s_2$ matches $s_1$. A well-trained network naturally realizes this via hopping through discrete attractors \cite{Rungratsameetaweemana2025}. After the first cue $s_1$, the network state settles near one of two stable fixed points $\mathbf{x}^*(s_1)$ (one for $s_1=+1$, one for $s_1=-1$), storing the first input in a sustained firing pattern. When the second cue $s_2$ arrives, it drives the network out of the $\mathbf{x}^*(s_1)$'s basin toward the new basin of a stable fixed point $\mathbf{x}^*(s_1,s_2)$. Because $s_2$ can also be $\pm1$, the second input causes a branching into one of four possible basins of the attractors $\{\mathbf{x}^*(+1,+1),\mathbf{x}^*(+1,-1),\mathbf{x}^*(-1,+1),\mathbf{x}^*(-1,-1)\}$, each encoding the ordered pair of cues (see state-space branching in Fig.~\ref{fig:wm_sketch}, bottom). Importantly, the final attractors are arranged in the space of states such that a simple hyperplane (linear readout) can separate the ``matched'' states ($s_1=s_2$) from the ``non-matched'' states ($s_1\neq s_2$). In this manner, the temporal logic of the task is implemented by the classification of the sustained firing pattern encoded in the eventual attractor $\mathbf{x}^*(s_1,s_2)$. Fig.~\ref{fig:wm_sketch} illustrates this attractor-based mechanism schematically for this two-step binary sequence memory.

A concern follows from this discrete-attractor picture: reliable sequence coding requires each cue to \emph{readily} push the state out of its current basin; thus these operating fixed points must be only \emph{weakly stable} (near-marginal), seemingly a fine-tuning challenge. 

\subsection{Neuronal networks with heterogeneous synaptic timescales trained for DMS tasks}\label{sec: neuronal networks}

It has been shown in Ref.~\cite{Rungratsameetaweemana2025} that the desirable landscape of \emph{near-marginal} discrete attractors emerges robustly when networks are trained with injected noise and the neuron-specific synaptic timescales are learnable. Concretely, \cite{Rungratsameetaweemana2025} employs a continuous-time firing-rate RNN with excitatory (E) and inhibitory (I) populations (Dale’s law; E/I ratio 80/20 mimicking cortical networks; $N=200$; network state $\mathbf{x}(t)\in\mathbb{R}^N$) and heterogeneous synaptic timescales $\{\tau_i\}$, described by the following dynamical model:
\begin{equation}
\label{eq:dyn_model}
\tau_i \dot x_i(t) = - x_i(t) + \sum_{j=1}^{N} W_{ij}\sigma\big(x_j(t)\big) + \sum_{k=1}^{U} W^{\mathrm{in}}_{ik}u_k(t) + \sum_{c=1}^{C} W^{\mathrm{noise}}_{ic}\xi_c(t),
\end{equation}
with sigmoidal firing rate $\sigma(x)=1/(1+e^{-x})$, sign-constrained weights $W_{ij}\ge0$ for $j\in E$ and $W_{ij}\le0$ for $j\in I$, inputs $u_k(t)$ providing the input cues, and $\xi_c(t)$ denoting independent Gaussian white-noise channels. A schematic depiction of the input and noise drives entering through $W^{\mathrm{in}}$ and $W^{\mathrm{noise}}$ is shown in Fig.~\ref{fig:wm_sketch}. A linear readout $o(t)=W^{\mathrm{out}}\sigma(\mathbf{x}(t))+b$ is taken from the firing-rate vector $\sigma(\mathbf{x}(t))$ during the post-$s_2$ response window. The network is trained on the DMS (temporal XNOR) task with two 250-ms stimuli $s_1,s_2\in\{-1,+1\}$ separated by a 250-ms delay, delivered through identical input channels $u_k(t)$; targets for the linear readout are $+1$ for matches and $-1$ for non-matches. Geometrically, the linear readout learns a hyperplane in the network firing-rate space that separates the match from the non-match classes. Trainable parameters are the recurrent weights $W$, the synaptic decay timescales $\{\tau_i\}$, the noise-input weights $W^{\mathrm{noise}}$, and the readout $(W^{\mathrm{out}},b)$. During testing, $W^{\mathrm{noise}}$ is replaced by a fresh Gaussian random matrix to assess the model robustness to noise (see details in Materials \& Methods in \cite{Rungratsameetaweemana2025}). 

\begin{figure}[t]
  \centering
  \includegraphics[width=.7\linewidth]{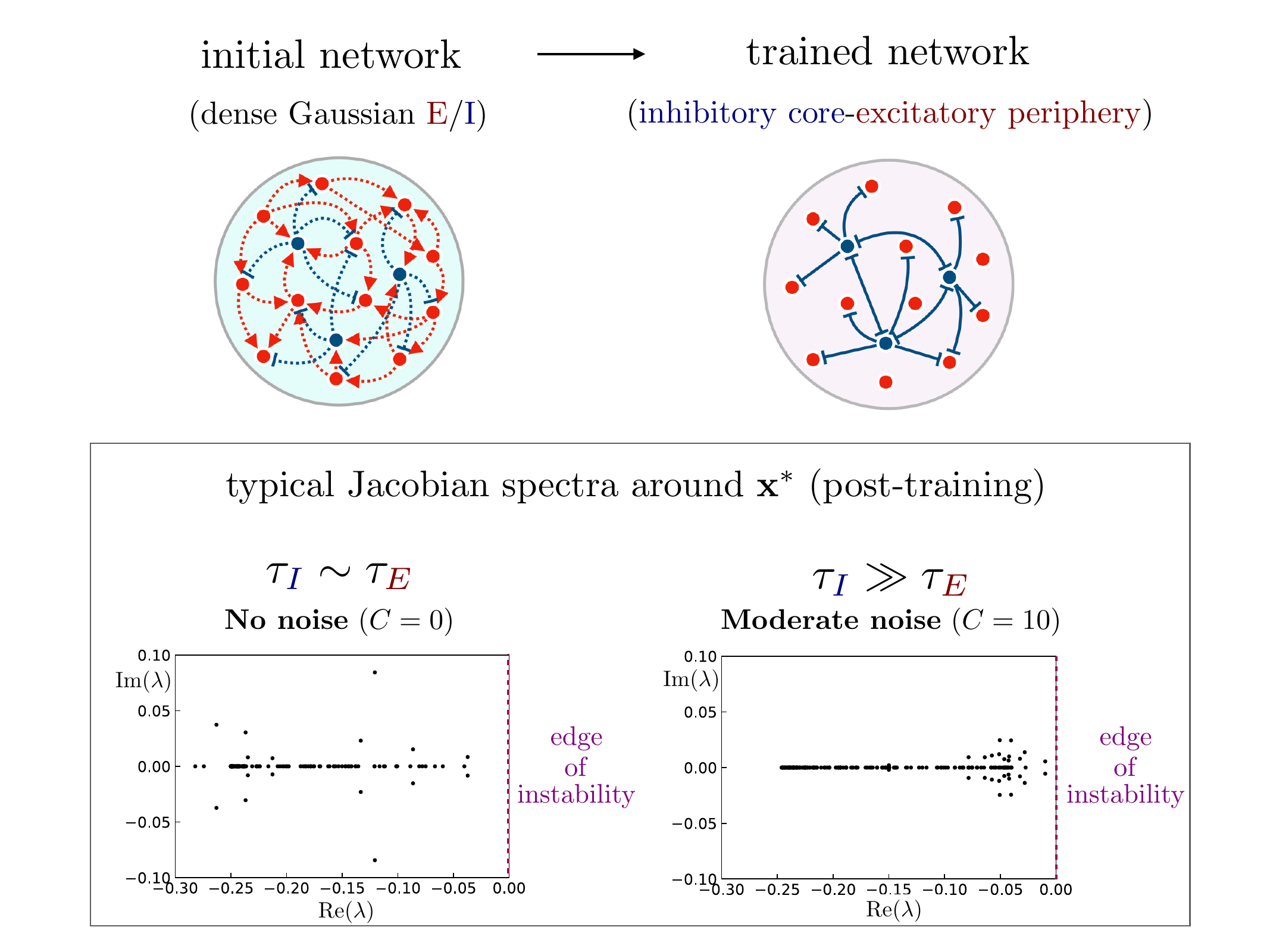}
 \caption{\textbf{Emergent inhibitory core---excitatory periphery motif and  typical Jacobian spectra after training with noise.}
Starting from a dense Gaussian E/I network, training on the DMS task with additive noise, as in Eq.~\eqref{eq:dyn_model}, yields an \emph{inhibitory core---excitatory periphery} architecture (top): excitatory outputs are pruned while inhibition dominates but relatively sparse. (Bottom): typical eigenvalue spectra of the Jacobian at a delay-period operating point $\mathbf{x}^*(s_1)$ of Fig.~\ref{fig:wm_sketch}. This Jacobian is characterized by the random matrix ensemble
\mbox{$J(\mathbf{x}^*)=\mathcal{T}^{-1}\left(-I+W\mathcal{H}^*\right)$}, Eqs.~(\ref{eq:Jacobian_operating_point}-\ref{eq:J_component}).
With weak-noise training (bottom left), $\tau_I \sim \tau_E$ and the bulk remains far from the edge-of-stability (vertical dashed line).
With moderate-noise training, inhibitory timescales separate by roughly an order of magnitude, producing a characteristic \emph{bar-and-blob} spectral geometry. As we will show, a real-axis bar is trivially set by excitatory timescales while slow inhibitory timescales push the inhibitory blob toward the edge-of-stability, i.e., near-marginal operating points (bottom right).
In Sec.~\ref{sec: setup_RMT} we formalize these post-training statistics with a sparse non-Hermitian E/I ensemble; Sec.~\ref{sec:jacobian-SUSY} develops an analytic random matrix theory that predicts the bar-and-blob shape and a spectral-edge condition linking sparsity, E/I ratio, and the distributions of $\tau$ and $h$ to the blob’s proximity to $\Re\lambda=0$.}
  \label{fig:motif_spectra}
\end{figure}

Increasing the number of independent noise channels (and thus effective variance) not only improves training success and test accuracy but also induces \emph{longer inhibitory} synaptic timescales. In addition, during the delay period, when the input drive is absent $u_k(t)=0$, the network state with slow inhibitory dynamics hovers near an operating attractor $\mathbf{x}^*(s_1)$, such that the corresponding delay-period Jacobian spectra around these operating attractors are \emph{shifted toward the stability boundary} while remaining stable, see Fig.~\ref{fig:motif_spectra} (bottom). This noise-shaped timescale heterogeneity yields operating fixed points that are generically \emph{near marginally stable} and readily driveable, exactly the desirable regime required for input-driven discrete attractor transitions. This motivates a random-matrix analysis of the Jacobians around these fixed points, the operating points labeled $\mathbf{x}^*(s_1)$ in Fig.~\ref{fig:wm_sketch}.

\subsection{Trained network characteristics and a sparse non-Hermitian random matrix ensemble}\label{sec: setup_RMT}

What network features yield \emph{near-marginal} operating-point discrete attractors? Empirically (see Appendix~\ref{app:network_stats}), trained DMS network weights exhibit an \emph{inhibitory core---excitatory periphery} motif: columns associated with excitatory neurons are strongly suppressed to 0 (excitatory units act largely as sinks), while columns associated with inhibitory neurons carry most of the recurrent feedback and are relatively sparse. This motivates the following statistical description of the Jacobians at delay-period operating fixed points.

\begin{figure}[h!]
    \centering
    \includegraphics[width=0.85\linewidth]{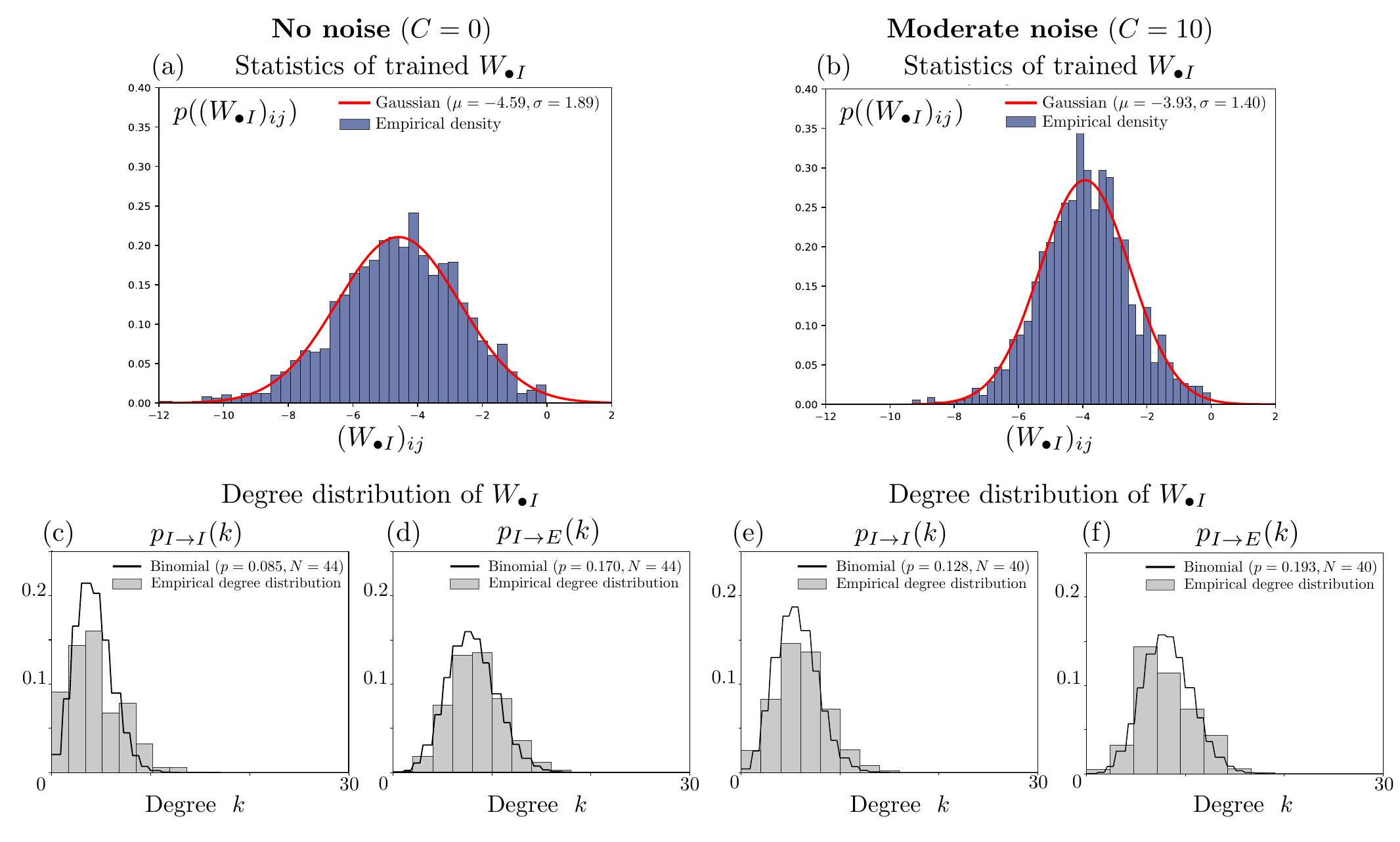}
    \caption{\textbf{Statistics of the trained inhibitory core.} 
Here, we characterize the distribution of the \emph{non-zero} inhibitory weights $W_{\bullet I}$ and their connectivity patterns, justifying the sparse non-Hermitian ensemble assumptions.
(a)-(b) The distribution of non-zero inhibitory weights fits a \emph{Gaussian distribution}.  
(c)-(f) The degree distributions (number of non-zero connections per neuron) for inhibitory-to-inhibitory ($I \to I$) and inhibitory-to-excitatory ($I \to E$) projections are well-captured by Binomial distributions. This empirically justifies modeling the connectivity matrix as a \emph{sparse Erd\H{o}s-R\'enyi graph} with \emph{finite} mean degree $k$, as defined in Eq.~\eqref{eq:sparse-ensemble}.
Statistics are derived from representative samples in the public dataset accompanying Ref.~\cite{Rungratsameetaweemana2025} (\url{https://osf.io/dqy3g/}).}
\label{fig:fitted_weight}
\end{figure}

Let $\mathbf{x}^*=\mathbf{x}^*(s_1)$ denote the attractor selected by the first cue. Linearizing \eqref{eq:dyn_model} about $\mathbf{x}^*$ during the delay gives
\begin{align}
  \frac{d}{dt}\delta \mathbf{x} &= J(\mathbf{x}^*)\delta \mathbf{x}, \qquad
  J(\mathbf{x}^*) = \mathcal{T}^{-1}\big(-I + W \mathcal{H}^*\big), \label{eq:Jacobian_operating_point}
\end{align}
with the \emph{learned timescale matrix} 
\begin{equation}
\mathcal{T}=\mathrm{diag}(\tau_1,\ldots,\tau_N),
\end{equation}
and the \emph{gain matrix}
\begin{align}
\mathcal{H}^* &=\mathrm{diag}(h_1,\ldots,h_N), \\
h_j &=\sigma'(x^*_j)=\sigma(x^*_j)\big(1-\sigma(x^*_j)\big)\in(0,1/4] \nonumber
\end{align}
are computed at the corresponding fixed point. 
Reordering indices into excitatory ($E$) and inhibitory ($I$) groups (rows: targets; columns: sources), the trained weight matrix decomposes as
\begin{equation}\label{eq: block_trained_weight}
  W =
  \begin{pmatrix}
     W_{EE} & W_{EI}\\
     W_{IE} & W_{II}
  \end{pmatrix}
  \approx
  \begin{pmatrix}
     0 & W_{EI}\\
     0 & W_{II}
  \end{pmatrix},
\end{equation}
i.e., excitatory-outgoing columns are suppressed ($W_{\bullet E}\approx0$), while inhibitory-outgoing columns ($W_{\bullet I}$) are well modeled as \emph{sparse} with connection probability $k_{\bullet}/N_I$ and block-dependent Gaussian inhibitory weights with moments $(\mu_{\bullet I},\sigma^2_{\bullet I})$, $\bullet\in\{E,I\}$ (see Fig.~\ref{fig:fitted_weight} and Appendix~\ref{app:network_stats}). Let $f =|I|/N$ denote the inhibitory fraction (empirically $f\approx0.2$ for cortical networks); because only the $fN$ inhibitory columns contribute substantially to recurrent feedback, the effective column subspace of $W$ is of size $fN$, giving the weight ensemble an \emph{effectively rectangular} column support, together with block structure and sparsity. 

\begin{figure}[h]
    \centering
    \includegraphics[width=0.7\linewidth]{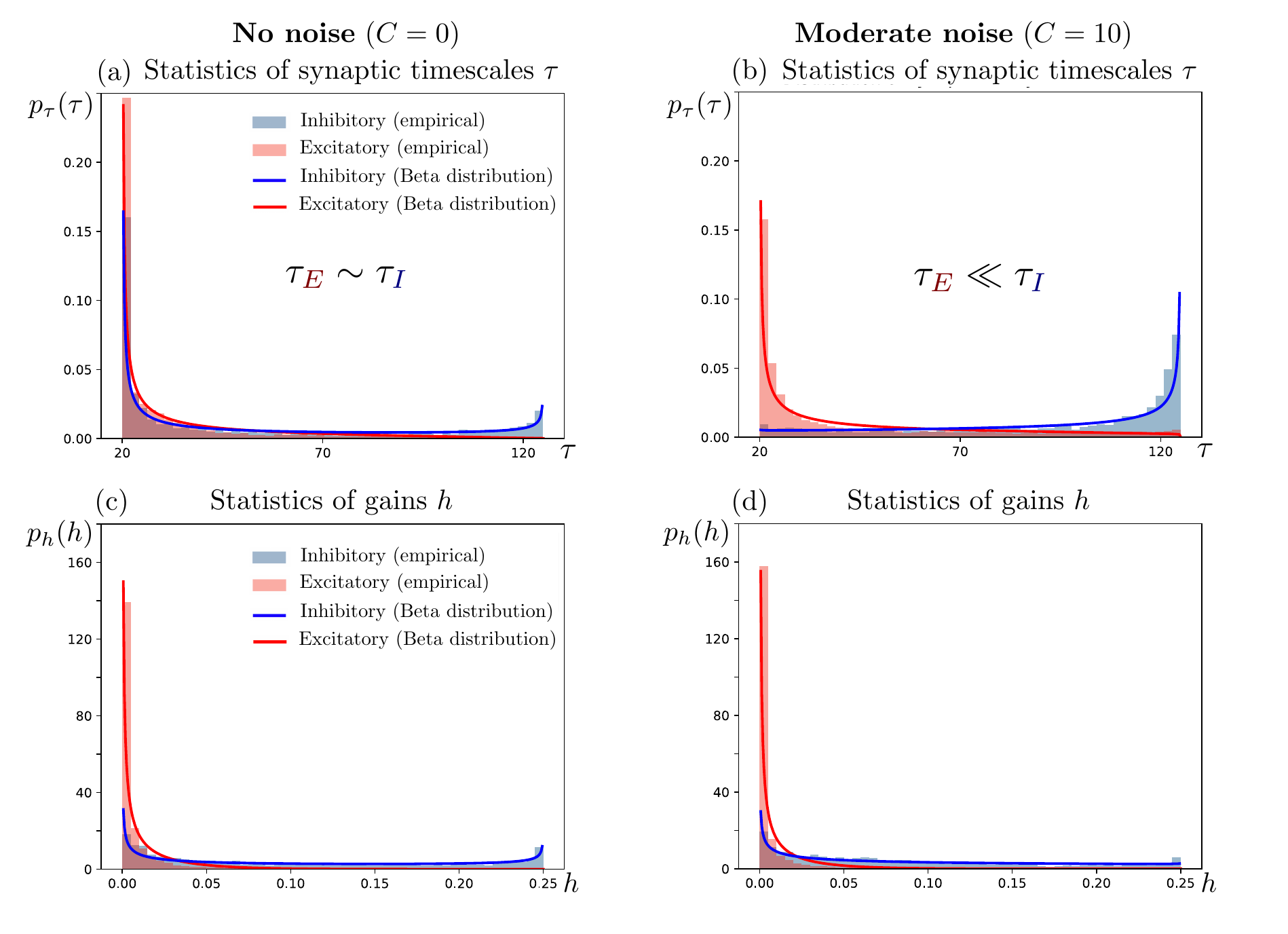}
    \caption{\textbf{Trained heterogeneous synaptic timescales and gain function distributions.} 
Histograms show the empirical statistics of synaptic decay timescales $\tau$ (top row) and activation gains $h = \sigma'(x^*) \in [0,1/4]$ (bottom row) for excitatory (red) and inhibitory (blue) populations, aggregated across all trained networks (50 samples). Solid lines are Beta distribution fits. The biological synaptic timescales from \cite{Rungratsameetaweemana2025} range from 25 to 125 ms.
(a) In networks trained without noise ($C=0$), both populations exhibit similar, fast timescale distributions ($\tau_E \sim \tau_I$). Fitted Beta parameters $(\alpha, \beta)$ are: Excitatory $(0.274, 1.66)$; Inhibitory $(0.273, 0.598)$.
(b) Training with moderate noise ($C=10$) induces a separation of timescales where inhibitory units develop \emph{significantly slower} dynamics ($\tau_E \ll \tau_I$). Fitted parameters: Excitatory $(0.35, 1.1)$; Inhibitory $(0.973, 0.473)$.
(c)-(d) The gain distributions remain sharply peaked near zero for both noise conditions.
(c) ($C=0$) Fitted parameters: Excitatory $(0.306, 6.51)$; Inhibitory $(0.496, 0.656)$.
(d) ($C=10$) Fitted parameters: Excitatory $(0.254, 6.88)$; Inhibitory $(0.565, 0.967)$.
}
\label{fig:fitted_tau_g}
\end{figure}

In component form, the Jacobian entries couple three distributions:
\begin{equation}
  J_{ij} = \frac{1}{\tau_i}\Big(-\delta_{ij} + W_{ij}h_j\Big).
  \label{eq:J_component}
\end{equation}
Namely, successfully trained models with moderate noise exhibit broad \emph{timescale heterogeneity} with inhibitory $\tau_i$ typically longer than excitatory, and both the $\{\tau_i\}$ and the gains $\{h_j\}$ are well fitted by bounded distributions---specifically, the properly scaled Beta distribution $x^{\alpha-1}(1-x)^{\beta-1}/B(\alpha,\beta)$, see Fig.~\ref{fig:fitted_tau_g}. 

These observations lead us to study the spectrum of the structured, sparse non-Hermitian Jacobian ensemble with sign constraints (\ref{eq:Jacobian_operating_point}-\ref{eq:J_component}) defined by
\begin{enumerate}
    \item  {\bf Sparse inhibitory core---excitatory periphery motif.}  Column-suppressed $W_{\bullet E} \approx 0$ and sparse inhibitory \emph{Gaussian} blocks with moments $(\mu_{\bullet I},\sigma^2_{\bullet I})$ and a \emph{finite} mean degree $k_{\bullet}$, corresponding to an  Erd\H{o}s-R\'enyi graph structure, see the definition of sparsity in Eq.~\eqref{eq:sparse-ensemble}.
    \item {\bf Heterogeneous synaptic timescales and gains statistics.} Diagonal heterogeneities $\mathcal{T}^{-1}$ (timescales) and $\mathcal{H}^*$ (gains) drawn from empirical Beta distributions.
    \item {\bf A tunable inhibitory fraction.} $f$ controlling the effective column support of $W$.
\end{enumerate}
A typical spectral geometry from the actual training is exemplified in Fig.~\ref{fig:motif_spectra} (bottom). Our goal is to determine the relation between these statistical properties and the placement of the spectral edge $\partial \mathrm{Spec}(J)$ close to $\Re\lambda=0$, thereby quantifying how near-marginal discrete attractors emerge.

\begin{figure}[t]
    \centering
    \includegraphics[width=0.8\linewidth]{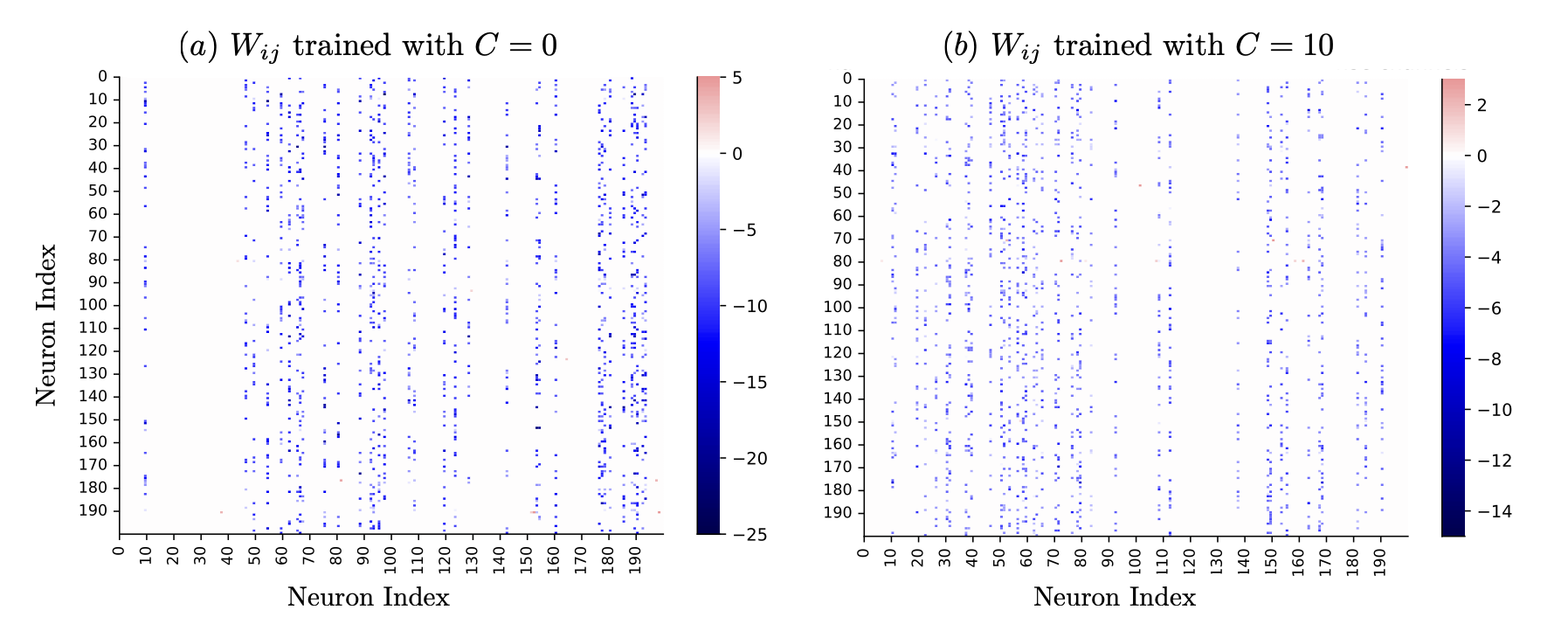}
    \caption{{\bf Typical trained synaptic weights}. The emergent $W_{ij}$ trained without noise $C=0$ (a) and with moderate noise $C=10$ (b) both exhibit inhibitory core---excitatory periphery motifs; outbound excitation is largely suppressed while inhibition dominates. The outgoing inhibitory weights are also quite sparse, see Fig. \ref{fig:fitted_weight} (bottom). Here $N = 200$. The displayed weights are representative samples obtained from the public dataset accompanying Ref. \cite{Rungratsameetaweemana2025} (\url{https://osf.io/dqy3g/}).}
    \label{fig:trained_weight}
\end{figure}

\paragraph*{Analytic strategy.}
The ensemble of interest is \emph{sparse, non-Hermitian, and block-structured}, so classical dense i.i.d. results (Ginibre/circular law) \cite{Ginibre1965,Girko1985} do not directly apply. Our starting point will be the \emph{Hermitized resolvent} representation for the eigenvalue density of a non-Hermitian matrix, to be introduced explicitly in the next section. This representation goes back in its basic form to \cite{Girko1985}, but provides an approach to much more sophisticated ensembles than that original context.
It was prominently revived in \cite{Feinberg1997} and explored in multiple subsequent works; see \cite{Metz2019} for a contemporary review. This representation
recasts the eigenvalue density of a non-Hermitian matrix through the inverse of a certain Hermitian matrix of doubled size. This matrix-inverse rewriting allows further processing using methods of quantum/statistical field theory, resulting in an explicit saddle point equation that controls the eigenvalue density in the limit of infinitely
large matrices. There are two dominant approaches to handling the treatments arising from this general idea: one based on integrals over bosonic replicas and one based
on supersymmetric integrals. We shall follow the latter approach and in particular its application to random matrices in a format that goes back to the works of Fyodorov and Mirlin \cite{Fyodorov1991, Mirlin_1991}; see \cite{Efetov1996,wegner} for textbook treatments and \cite{Akarapipattana2023Random} for a contemporary pedagogical exposition for the case of Hermitian matrices. The ensemble (\ref{eq:J_component}) is complicated because of its combination of sparsity, block structure, non-Hermiticity, and modification of the random matrix $W$ with extra random factors $\tau_i$ and $h_j$. This is considerably outside what has been treated in the past random matrix literature, but different components of the problem have been considered before. For instance, one can see a treatment of non-sparse Hermitian block random matrices in \cite{block}. Sparse non-Hermitian matrices without block structure and modifications with synaptic timescales and gains have been considered in \cite{Metz2019,Baron2023,sparse2025}, considerations of sparse Hermitian matrices using the supersymmetry-based approach can be seen in \cite{Fyodorov1991, Mirlin_1991,Akarapipattana2023Random,Akarapipattana2025Hammerstein,Akarapipattana2025Statistical},
while some other related non-sparse non-Hermitian ensembles have been considered in \cite{lindbladian,spindle,hebbian}.
We will have to combine these diverse ingredients into an effective computational scheme to analyze the ensemble (\ref{eq:J_component}), which will be done in Section~\ref{sec:jacobian-SUSY}. Before that, we shall however describe how a more straightforward ensemble of sparse non-Hermitian matrices is treated using the methods we have in mind. This will provide a pedagogical introduction before turning to the more sophisticated computations for the ensemble (\ref{eq:J_component}), while the integral saddle point equation that we shall derive has not appeared in the literature before to the best of our knowledge and is of value in itself from the standpoint of random matrix theory. 

\section{Statistical field theory of sparse non-Hermitian random matrices}
\label{sec:intro_susy}

We proceed to develop, step by step, a statistical field theory approach to analytically calculate the spectral density
of a broad class of \emph{sparse, non-Hermitian} random matrices. Our goal is to introduce, in a pedagogical manner, the key techniques (Hermitization, supersymmetric Gaussian integrals, and the Fyodorov-Mirlin decoupling) before moving to the biologically
motivated Jacobians from Sec.~\ref{sec: setup_RMT} in the next section. For the illustration in this section, we choose the simplest sparse non-Hermitian ensemble of $N\times N$ matrices with, on average, $k$ nonzero entries per row, treated in the limit where $N$ goes to $\infty$ and $k$ stays fixed.

Readers unfamiliar with supersymmetry (SUSY) in application to statistical physics can treat it as a useful mathematical device for analytically computing the ensemble average of the resolvent: it converts matrix inverses of the form $(M-zI)^{-1}$, which are difficult to average directly, into Gaussian integrals. This trick moves the random matrix $M$ from the denominator into the exponent, where the ensemble average $\mathbb{E}[\cdot]$ can then be computed easily in terms of characteristic functions
of the probability distributions of the matrix entries. 
This SUSY integration requires a set of auxiliary variables: standard commuting variables (bosons) and anticommuting Grassmann variables (fermions). The latter are a convenient mathematical bookkeeping device, which ensures in an economical manner that the entire integral is mathematically identical to the matrix inverse we began with.

\subsection{Ensemble of interest and its Hermitized representation}\label{subsec:hermitization}

Let $M\in\mathbb{R}^{N\times N}$ be a random \emph{real} non-Hermitian matrix with i.i.d. entries drawn from a
\emph{sparse} ensemble
\begin{equation}
  \label{eq:sparse-ensemble}
  \mathbb{P}(M) = \prod_{a,b=1}^N
  \Bigg[\Big(1-\frac{k}{N}\Big)\,\delta(M_{ab}) + \frac{k}{N}\,p(M_{ab})\Bigg],
\end{equation}
where $k>0$ is the finite ($N$-independent) mean degree and $p(\cdot)$ is a probability density for the
nonzero matrix entries. 

The ensemble-averaged spectral density of $M$ in the complex plane is defined  --- see \cite{Metz2019}, for example  --- by 
\begin{equation}
  \label{eq:girko}
  \rho(z) = \frac{1}{\pi}\,\partial_{\bar z} G(z),
  \qquad
 G(z)=-\frac1N\mathbb{E}\Tr(M-zI)^{-1}
\end{equation}
where $z=x+iy\in\mathbb{C}$, $\partial_{\bar z}=\frac{1}{2}(\partial_x + i\partial_y)$, and
$\mathbb{E}$ denotes ensemble average over $M$. Indeed, if the eigenvalues of $M$ are $z_i$, $\Tr(M-zI)^{-1}=\sum_i 1/(z_i-z)$. Furthermore,
\beq\label{electrostat}
\del_{\bar z}\,\frac1{z_i-z}=-\,\del_{\bar z}\del_z \log(z_i-z)=-\pi\de(z-z_i),
\eeq
where $\de(z)$ is the two-dimensional $\delta$-function in the complex plane, recovering the definition 
\beq\label{rhodef}
\rho(z)=\frac1N\sum_i \de(z-z_i). 
\eeq
Note that (\ref{electrostat})
is directly related to the two-dimensional electrostatic potential of a point charge, expressed in the complex plane notation. (A more direct way to understand (\ref{electrostat}) is by writing $\del_{\bar z}[1/z]=\lim_{\eps\to 0}\del_{\bar z}[\bar z/(\eps^2+z\bar z)]=\lim_{\eps\to 0}[\eps^2/(\eps^2+z\bar z)^2]=\pi\de(z)$. The last step
follows from the observation that the limit is 0 for any $z\ne 0$, while the integral over the $z$-plane is $\pi$, independent of $\eps$.)

Representing the eigenvalue density through the matrix inverse appears more manageable than writing an explicit sum over the eigenvalues in the definition (\ref{rhodef}), but it still cannot be averaged directly over the ensemble (\ref{eq:sparse-ensemble}). The next point in the strategy is to develop representations of  (\ref{eq:girko}) in terms of Gaussian integrals that will make it possible to average over the ensemble (\ref{eq:sparse-ensemble}). To get an idea why this may be possible (the precise implementation will follow below), consider the following two standard multidimensional Gaussian integration formulas: take an $N\times N$ matrix $A$ and write:
\begin{align}
&[A^{-1}]_{kl}=\det A\, \int [dx]\, \bar x_l x_k \exp\left[-\sum_{lm} A_{lm}\bar x_l x_m\right],\label{Gaussboson1}\\
&[A^{-1}]_{kl}=\det[-iA]\, \int [dx]  (-i\bar x_l x_k) \exp\left[i\sum_{lm} A_{lm}\bar x_l x_m\right]\label{Gaussboson2}.
\end{align}
The integration is over $N$-dimensional complex vectors $(x_1,x_2,\cdots,x_N)$ so that 
\beq
[dx]\equiv \prod_{k=1}^N\frac{d\bar x_k dx_k}{\pi},\qquad d\bar z dz\equiv d\Re(z)d\Im(z).
\eeq
Both formulas are correct, but require different conditions on $A$ to ensure convergence of the integrals on the right-hand side. 
Formula (\ref{Gaussboson1}) works if all eigenvalues of $A$ have positive real parts. It is normally not usable for random $A$ unless the ensemble is such that this condition is automatically satisfied. Formula (\ref{Gaussboson2}) works for any $A$ whose eigenvalues have nonnegative imaginary parts. In particular, it works for real eigenvalues, and hence, for any Hermitian $A$. With some adaptations, we will be able to use this formula for our purposes. (For a Hermitian $A$, the above formulas are easily proved by rotating to the eigenbasis of $A$, whereupon the integral factorizes into independent  integrals over one complex variable each.)

There are two obstructions for immediately applying (\ref{Gaussboson2}) to simplify the averaging of (\ref{eq:girko}). First, the right-hand sides in (\ref{Gaussboson1}) and (\ref{Gaussboson2}) contain determinant factors that are complicated functions of $A$ that cannot be averaged immediately (by contrast, the exponential in the integrand is conveniently factorized over the entries of $A$). This will be tackled below by representing this determinant as a Gaussian integral over {\it anticommuting} variables, which will make it possible to average the entire expression easily over the ensemble (\ref{eq:sparse-ensemble}). The second problem is that the matrix whose inverse we need to average, $M-zI$, is non-Hermitian. This point will  be dealt with immediately by an appeal to the {\it Hermitization method} \cite{Feinberg1997, Girko1985, Metz2019}.

The strategy is to replace the $N\times N$ non-Hermitian matrix $M-zI$, whose inverse we must compute, with the following $2N\times 2N$ Hermitian matrix:
\begin{equation}
  H(z)=
  \begin{pmatrix}
   0& -i(M-zI) \\
    i(M^{\dagger}-\bar z I) & 0
  \end{pmatrix}.
  \label{eq:herm-Heps}
\end{equation}
By construction, $H(z)^\dagger=H(z)$. It can be verified by a direct computation that the inverse of this matrix is
\begin{equation}
  \label{eq:hermitization-inverse}
  H(z)^{-1}
  =
  \begin{pmatrix}
    0 & -i(M^\dagger-\bar z I)^{-1}\\[2pt]
    i(M-zI)^{-1} & 0
  \end{pmatrix},
\end{equation}
and it contains in the lower-left block the resolvent  (\ref{eq:girko}). Thus, to compute the resolvent  (\ref{eq:girko}), we can appeal to the inverse of the Hermitian
matrix $H(z)$, and in this case a convergent Gaussian representation (\ref{Gaussboson2}) is available. It remains to explain how to deal with the remaining determinant factor.

\subsection{Supersymmetric Gaussian integral for the resolvent and ensemble averaging}\label{secGauss}

By linearity of the trace and the expectation, the averaged normalized resolvent is $G(z) = -\frac{1}{N}\mathbb{E}\Tr(M-zI)^{-1} = -\frac{1}{N}\sum_{k=1}^N \mathbb{E}\big[(M-zI)^{-1}_{kk}\big]$. For the i.i.d. sparse ensemble defined in \eqref{eq:sparse-ensemble}, every index is statistically identical. This homogeneity implies that the average of any diagonal element is the same as any other, so $\mathbb{E}\big[(M-zI)^{-1}_{11}\big] = \mathbb{E}\big[(M-zI)^{-1}_{kk}\big]$ for all $k$. The sum therefore simplifies to 
\begin{equation}
    G(z) =- \frac{1}{N} \sum_{k=1}^N \mathbb{E}\big[(M-zI)^{-1}_{11}\big] =- \frac{1}{N} \Big( N \cdot \mathbb{E}\big[(M-zI)^{-1}_{11}\big] \Big) = -\mathbb{E}\big[(M-zI)^{-1}_{11}\big].
\end{equation}
Thus, the problem of computing the full averaged trace reduces to computing the average of a single, arbitrary diagonal element, which we take to be the $(1,1)$ element.

To represent $(M-zI)^{-1}_{11}$ as a Gaussian integral we then employ  (\ref{Gaussboson2}) for the Hermitian matrix $H(z)$ defined by (\ref{eq:herm-Heps}). Since $H(z)$ is a $2N\times 2N$ matrix, we need a complex vector with $2N$ components to perform the Gaussian integral. This vector is most conveniently represented as a pair of vectors with $N$ components each that we denote $x$ and $\tilde x$. With this, together with  (\ref{Gaussboson2}) and (\ref{eq:herm-Heps}), we write:
\beq
-(M-zI)^{-1}_{11}=\det[-iH(z)]\, \int [dx][d\tilde x] \,(\bar x_1 \tilde x_1) \exp\left\{\sum_{a,b=1}^N \Big[ \bar x_a (M_{ab} - z\delta_{ab}) \tilde x_b - \bar{\tilde x}_b (M_{ab}- \bar z \delta_{ab} )x_a\Big]\right\}\label{bosonresolv}.
\eeq

It remains to express $\det[H(z)]$ in a way factorized over the entries of $M$, and we employ integrals over anticommuting (Grassmannian) variables for this purpose.
Such representations of determinants are common in quantum field theory (and have been prominently used in the Faddeev-Popov path-integral quantization of gauge theories). We define anticommuting variables $\xi_a$ by the relations
\beq\label{xidef}
\xi_a \xi_b=-\xi_b\xi_a.
\eeq 
We will use the complex Grassmannian notation in which the `complex conjugates' $\bar\xi_a$ are independent of $\xi_a$ and they anticommute with all $\xi$'s and among themselves.

The anticommutation relations imply, importantly, that
\beq\label{xisq}
\forall a:\qquad \xi_a^2=0.
\eeq
This means that the algebra of functions of $\xi$ is extremely simple: all functions are at most linear in each $\xi_a$. There is a standard specification of integration rules on such linear functions of $\xi$, originally due to Berezin, that says:
\beq\label{Berezin}
\int d\xi\,(b+a\xi)=a.
\eeq
In other words, a constant integrates to 0, while $\xi$ integrates to 1. (If there are multiple $\xi$-variables, as in our case, they should be reordered before applying the above formula so that the variable that is being integrated stands next to its own differential, while all the differentials are treated as anticommuting, similar to the variables themselves. This fixes unambiguously the $\pm$ signs.) Note that one should not think of these integrals as Riemann-type sums. They are explicitly defined maps from functions of $\xi$ to numbers that share properties of ordinary integrals and have a number of useful features. To clarify, in application to a function of $\xi$ and $\bar\xi$, we have
\beq
\int d\bar\xi d\xi \,(d+c\xi+b\bar\xi+a\bar\xi \xi)=-a.
\eeq
The most useful property in our context is
the Gaussian representation for determinants:
\beq\label{detBerezin}
\det A=\int[d\xi] \exp\left[-\sum_{a,b=1}^N A_{ab}\bar\xi_a \xi_b\right],
\eeq
with $[d\xi]\equiv \prod_n d\bar\xi_n d\xi_n$.
This formula holds for any matrix $A$ without restrictions, as issues of convergence do not arise for Grassmannian integrals. To prove this formula, one expands the exponential in a Taylor series, which will truncate at a finite number of terms due to (\ref{xisq}). Then, (\ref{xisq}) will eliminate all terms except for those where each $\xi_a$ and each $\bar\xi_a$ is present exactly once. Each such term will come with a product of exactly $N$ entries of $A$, while the $\pm$ signs generated by permuting the anticommuting variables in the course of integration will precisely supply the permutation parity factors necessary for assembling the correct formula for the determinant.

We then apply (\ref{detBerezin}) with $-iH(z)$ substituted for $A$ to represent the $\det[-iH(z)]$ factor in (\ref{bosonresolv}). Since $H(z)$ is a $2N\times 2N$ matrix, we have to apply the doubling to the anticommuting variables, resulting in two sets: $\xi_a$ and $\tilde\xi_a$. It is convenient to bundle the bosonic variables $x_a$ and $\tilde x_a$ in (\ref{bosonresolv}) with these newly introduced fermionic variables to obtain the \emph{supervectors} $X_a = (x_a,\xi_a)$ and $\tilde X_a=(\tilde x_a,\tilde\xi_a)$. Then, all the ingredients can be assembled into the following {\it supervector} representation of the resolvent:
\begin{equation}
\label{eq:resolvent-elem}
-(M-zI)^{-1}_{11}
=
\int [dX]\,[d\tilde X]\,
\bar x_1\tilde x_1\prod_{a,b=1}^N
\exp \Big[X_a^\dag (M_{ab}- z\delta_{ab}) \tilde X_b - \tilde X_b^\dag (M_{ab} - \bar z \delta_{ab}) X_a\Big],
\end{equation}
where the integration measures are defined as
\beq
[dX]\equiv \prod_{a=1}^N\frac{d\bar x_a dx_ad\bar\xi_ad\xi_a}{\pi},\qquad [d\tilde X]\equiv \prod_{a=1}^N\frac{d\bar{\tilde x}_a d\tilde x_ad\bar{\tilde\xi}_ad\tilde\xi_a}{\pi},
\eeq
and the `scalar products' of supervectors are defined as
\beq
X_1^\dag X_2=\bar x_1 x_2+\bar\xi_1 \xi_2.
\eeq
If one unpackages (\ref{eq:resolvent-elem}) using these definitions, the integrals over $(x,\tilde x)$ and $(\xi,\tilde\xi)$ factorize, the latter produces a factor of  $\det[-iH(z)]$ by (\ref{detBerezin}), and then the former recovers the resolvent by (\ref{bosonresolv}). (Note that the complex conjugation of products of Grassmannian variables is defined to reverse the multiplication order, so that $X_1^\dag X_2=\overline{X_2^\dag X_1}$.)

An advantage of (\ref{eq:resolvent-elem}) is that, now, the dependences on the matrix entries of $M$ completely factorize, and the averaging over the ensemble
\eqref{eq:sparse-ensemble} is performed straightforwardly entry-by-entry
(for a derivation, see Appendix~\ref{app:susy}), yielding
\begin{align}
  \label{eq:after-average}
G(z)
  =
  \int [dX]\,[d\tilde X]\,
\bar x_1\tilde x_1
  \exp\Big[
    -\frac{1}{N}\sum_{a,b=1}^N C(X_a,\tilde X_b)
    -\sum_{a=1}^N\big(zX_a^\dag \tilde X_a - \bar z\tilde X_a^\dag X_a\big)
  \Big],
\end{align}
where the \emph{interaction kernel} $C$ encodes the \emph{weight distribution} $p(w)$ appearing in (\ref{eq:sparse-ensemble}) via its
characteristic function:
\begin{equation}
  \label{eq:C-kernel}
  C(X,\tilde X) = k\left[1-\hat p\big(i\tilde X^\dag X-iX^\dag \tilde X\big) \right],
  \qquad
  \hat p(t) = \int_{\mathbb{R}} e^{itw}p(w)dw.
\end{equation}
The integral in (\ref{eq:after-average}) is of the form that may be referred to as a large $N$ {\it supervector model} in the statistical field theory literature, and it remains
to solve it in the limit where $N$ becomes large.

\subsection{Fyodorov-Mirlin decoupling and the saddle-point equation}

The key obstacle for evaluating \eqref{eq:after-average} is the pairwise interaction
$\sum_{a,b}C(X_a,\tilde X_b)$, which couples the variables on all sites. An elegant approach to effectively bypassing this issue goes back to Fyodorov and Mirlin \cite{Fyodorov1991,Mirlin_1991} and relies on a functional analog of the Hubbard-Stratonovich transformation. In practice, for the purpose of our current
derivation we introduce a complex-valued function $q$ of a supervector argument and write:
\begin{equation}
  \label{eq:FM-identity}
  e^{-\sum_{ab}C(X_a,\tilde X_b)/N}
  =
  \int \mathcal{D}\bar q\,\mathcal{D}q\,\,
  \exp\!
    \int dX \left[-N\bar q(X)q(X)
   +i\sum_{b=1}^N \bar q(X)\,Q(X,\tilde X_b) +i\sum_{a=1}^N q(X)\,Q(X_a,X)\right],
\end{equation}
where $Q$ is the (operator) square root of $C$ satisfying
\beq
\int d\tilde X \,Q(X,\tilde X)\, Q(\tilde X,X')=C(X,X').
\eeq
We will not need the explicit form of $Q$ in practice since the final results will be re-expressed completely through $C$. The decoupling relation (\ref{eq:FM-identity}) is
nothing but the functional analog of the elementary one-dimensional complex Gaussian integration formula
\beq
\int\frac{d\bar z dz}{\pi}\, e^{-\bar z z +az+b\bar z}=e^{ab}.
\eeq

Substituting \eqref{eq:FM-identity} into \eqref{eq:after-average}, the integral over
$\{X_a,\tilde X_a\}_{a=1}^N$ \emph{factorizes} into $N$ identical copies  (except that the first one is modified by the {\it insertion} $\bar x_1\tilde x_1$), and one obtains
\begin{align}
  \label{eq:resolvent-factored}
 G(z)
  &=
  \int \mathcal{D}\bar q\,\mathcal{D}q \,\,\, \frac{ \int dXd\tilde X\,\bar x\tilde x\,\exp\left\{i\int dX' \left[\bar q(X')Q(X',\tilde X)+q(X')Q(X,X')\right] - z X^\dag \tilde X + \bar z\tilde X^\dag X\right\}}
        {\int dXd\tilde X\,\exp\left\{i\int dX' \left[\bar q(X')Q(X',\tilde X)+q(X')Q(X,X')\right] - z X^\dag \tilde X + \bar z\tilde X^\dag X\right\}}\,e^{-NS[q,\bar q]},
  \\
  S[q,\bar q] &\equiv
  \int dX\, \bar q(X)q(X)
  -\log\left(\int dXd\tilde X\,\exp\left\{i\int dX' \left[\bar q(X')Q(X',\tilde X)+q(X')Q(X,X')\right] - z X^\dag \tilde X + \bar z\tilde X^\dag X\right\}\right).\nonumber
\end{align}
In the large-$N$ limit, the functional integral over $q$ is dominated by
saddle points of the effective action $\delta S/\delta \bar q=\delta S/\delta q=0$. This reduces the problem to 
self-consistent saddle-point equations.

Evaluating the functional derivative $\delta S/\delta \bar q$ yields:
\begin{align}
\label{eq:saddle_q_raw}
q(X) =\, i\,\,\frac{ \displaystyle \int dX' d\tilde X \ Q(X,\tilde X)\, \exp\left\{i\int \!dY \!\left[\bar q(Y)Q(Y,\tilde X)+q(Y)Q(X',Y)\right] - z X'^\dag \tilde X + \bar z\tilde X^\dag X'\right\} }
 {\displaystyle\int dX'd\tilde X\,\exp\left\{i\int\! dY\! \left[\bar q(Y)Q(Y,\tilde X)+q(Y)Q(X',Y)\right] - z X'^\dag \tilde X + \bar z\tilde X^\dag X'\right\}},
\end{align}
while $\delta S/\delta  q$ produces the complex conjugate of this equation.
It is convenient to recast (\ref{eq:saddle_q_raw}) in terms of
\beq\label{qg}
g(X)\equiv \int dY \,q(Y)\,Q(X,Y),\qquad \tilde g(X)\equiv \int dY\,\bar q(Y)\,Q(Y,X),
\eeq
which lets one eliminate $Q$ in favor of $C$, writing the saddle point equations more compactly as:
\beq
\label{eq:saddle_g_raw1}
g(X) =\,i \,\,\frac{ \displaystyle \int dX' d\tilde X \, C(X,\tilde X)\, e^{i\tilde g(\tilde X)+ig(X') - z X'^\dag \tilde X + \bar z\tilde X^\dag X'} }
{ \displaystyle \int dX' d\tilde X\, e^{i\tilde g(\tilde X)+ig(X') - z X'^\dag \tilde X + \bar z\tilde X^\dag X'} }.
\eeq
This equation must be supplemented with the relation between $\tilde g$ and $\bar g$ that follows from the definitions (\ref{qg}):
\beq\label{gtildegbar}
\tilde g=\sign(\hat C) \,\bar g\equiv \int dX\, S(X,X')\,\bar g(X'),
\eeq
where $\sign(\hat C)$ is the operator, whose kernel we denote $S(X,X')$, derived by applying the sign function to $\hat C$ defined by $[\hat C g](X)=\int dX'\,C(X,X') \,g(X')$, so that the eigenfunctions of 
$\sign(\hat C)$ are the same as the eigenfunctions of $\hat C$ while their eigenvalues are $\,\pm 1$ depending on whether the corresponding eigenvalues of $\hat C$ are positive or negative. (Note that $\hat C$ is formally Hermitian as $C(X,X')=\overline{C(X',X)}$.) Some discussion of spectral properties of operators similar to $\hat C$ (in the context of Hermitian matrices) can be found at the end of section 2 of \cite{Mirlin_1991}, and also in the mathematical literature on `supersymmetric transfer operators' \cite{transfer1, transfer2}.

The saddle point equation (\ref{eq:saddle_g_raw1}) is invariant under {\it superrotations}: those are the linear transformations of $X$ that preserve the scalar product $X^\dag X'$. Such superrotations play the same role for supervectors as ordinary rotations for ordinary vectors. If $g(X)$ satisfies  (\ref{eq:saddle_g_raw1}), so does $g$ with a superrotated argument. This leaves two options for the solutions of (\ref{eq:saddle_g_raw1}): either $g$ is itself invariant under such superrotations so that no new solutions are produced when a superrotation is applied to $g$, or there is a whole family of saddles connected to each other by superrotations. It is customary to assume that $g$ is invariant, meaning that it only depends on the scalar product $X^\dag X$:
\beq\label{ginv}
g(X)=g(X^\dag X),
\eeq
and explore the consequences. As commonly seen in the literature on supersymmetry-based solutions for random matrices \cite{Fyodorov1991,Mirlin_1991,Akarapipattana2023Random}, this will correctly capture the dominant saddle point.

Under the assumption (\ref{ginv}), the equations simplify. For any integrable $F$, one has the following identities \cite{Efetov1996}:
\begin{equation}
  \int dX\,F(X^\dag X)=F(0),\qquad
  \int dXd\tilde X\,F(X^\dag X,\tilde X^\dag \tilde X,X^\dag \tilde X)=F(0,0,0).
  \label{eq:pf-efetov}
\end{equation}
Using \eqref{eq:pf-efetov} with $F=e^{i\tilde g(\tilde X)+ig(X') - z X'^\dag \tilde X + \bar z\tilde X^\dag X'}$ and keeping in mind that 
\beq
g(0)=0
\eeq 
since $C(0,\tilde X)=C(X',0)=0$,
the denominator of (\ref{eq:saddle_g_raw1}) equals $1$. Hence
\begin{equation}
  \label{eq:pf-saddle-no-den}
g(X)= i  \int dX' d\tilde X \, C(X,\tilde X)\, e^{i\tilde g(\tilde X)+ig(X') - z X'^\dag \tilde X + \bar z\tilde X^\dag X'}.
\end{equation}
In fact, it is possible to simplify this equation further and recast it as an ordinary one-dimensional integral equation for $g$ as a function of one variable with a Bessel kernel. 
This representation may have important applications, for example, for solving such equations numerically, but it will not play any direct role in our considerations here that focus mostly on the shape of the spectral edge. We summarize the relevant derivations in  Appendix~\ref{app:proof-rotational-invariant-saddle}.

If a solution of (\ref{eq:saddle_g_raw1}) has been obtained,
we can recover a saddle point estimate of $G(z)$ from (\ref{eq:resolvent-factored}). In this estimate, the saddle point value of the action $S[q,\bar q]$ plays no role due to the supersymmetry condition (\ref{ginv}) --- indeed, if $S$ evaluated at the saddle were nonzero, the large $N$ limit of the eigenvalue density would not exist. What remains is the $\bar x\tilde x$-insertion, evaluated on the solution of (\ref{eq:pf-saddle-no-den}), while the denominator equals 1 under the assumption (\ref{ginv}) by the same argument as above, leaving
\beq\label{Gg}
G(z)=
 \displaystyle \int dX d\tilde X\,\bar x\tilde x \,e^{i\tilde g(\tilde X)+ig(X) - z X^\dag \tilde X + \bar z\tilde X^\dag X}.
\eeq
Once again, this representation can be recast as an ordinary one-dimensional integral with a Bessel kernel, as we show in Appendix~\ref{app:proof-rotational-invariant-G}, but we will not need those formulas for our immediate purposes here.
Finally, from $G(z)$, one can recover $\rho(z)$ using (\ref{eq:girko}).

\subsection{Low sparsity limit and the circular law}
\label{subsec:saddle_to_spectrum}

To get a sense of how equation (\ref{eq:pf-saddle-no-den}) works, it is instructive to consider the limit of large $k$, which leads to an exact solution. Note that having a large $k$ does not mean that the matrix is dense: $k$ is still much smaller than $N$ (the latter already taken to infinity in the saddle point equation). However, as $k$ increases, even if it stays much smaller than $N$, the eigenvalue density approaches in shape that of dense non-Hermitian matrices, described by the circular law. (For a comprehensive contemporary discussion of the circular law from a mathematical perspective, see \cite{around}.)

To show how this limit is accommodated within our formalism, we restore $C$ of (\ref{eq:C-kernel}) explicitly in (\ref{eq:saddle_g_raw1}) to write
\begin{equation}\label{sddlkexp}
g(X)=  i k \int_{\mathbb{R}} dw\, p(w)\int dX' d\tilde X \left[1-e^{wX^\dag \tilde X-w\tilde X^\dag X} \right]\, e^{i\tilde g(\tilde X)+ig(X') - z X'^\dag \tilde X + \bar z\tilde X^\dag X'}.
\end{equation}
When $k$ is large, $g$ is of order $k$ and the integrals on the right-hand side acquire a saddle point structure due to a large argument of the exponential. We furthermore know that $g(X)=g(X^\dag X)$ and $g(0)=0$. We then expand $g$ and $\tilde g$ as
\beq\label{ggtldexp}
g(X)=k\left(uX^\dag X+\sum_{n=2}^\infty u_n (X^\dag X)^n\right),\qquad \tilde g(X)=k\left(\tilde uX^\dag X+\sum_{n=2}^\infty \tilde u_n (X^\dag X)^n\right).
\eeq
Substitution of (\ref{ggtldexp}) into (\ref{sddlkexp}) leads to the usual saddle point structure: the integral is dominated by small $\tilde X$ of order $1/\sqrt{k}$, the higher order corrections to $g$ controlled by $u_n$ become irrelevant at leading order in $1/k$, while the content of the square brackets in (\ref{sddlkexp}) can be Taylor-expanded as
\beq
1-e^{wX^\dag \tilde X-w\tilde X^\dag X} =-w\left(X^\dag \tilde X-\tilde X^\dag X\right)-\frac{w^2}2\left(X^\dag \tilde X-\tilde X^\dag X\right)^2+\cdots.
\eeq
The linear term does not have enough copies of $\tilde\xi$ to survive the fermionic integration over $\tilde\xi$ and hence does not contribute. One is then left with
\begin{equation}\label{sddldns}
g(X)=  -\frac{ik\sigma^2}2 \int dX' d\tilde X \left(X^\dag \tilde X-\tilde X^\dag X\right)^2\, e^{i\tilde g(\tilde X)+ig(X') - z X'^\dag \tilde X + \bar z\tilde X^\dag X'},
\end{equation}
with
\beq
\sigma^2\equiv  \int_{\mathbb{R}} dw\, p(w)\,w^2,
\eeq
which has to be solved with the self-consistent ansatz
\beq\label{ggtldtrunc}
g(X)=k\,u\,X^\dag X,\qquad \tilde g(X)=k\,\tilde u\,X^\dag X.
\eeq
All the integrals in (\ref{sddldns}) are super-Gaussian, and can be evaluated using formulas like (\ref{Gaussboson1}), (\ref{Gaussboson2}) and (\ref{detBerezin}). Since $g(X)$ only depends on the combination $X^\dag X$, complete information is obtained by evaluating it on a supervector $X=(x,0)$ with a real $x$ and a vanishing fermionic part. Thus,
$$
g(X)=g(X^\dag X)=g(x^2)= -\frac{ik\sigma^2x^2}2 \int d\bar x'dx'd\bar\xi'd\xi' d\bar{\tilde x}d\tilde xd\bar{\tilde\xi }d\tilde \xi \left(\tilde x-\bar{\tilde x}\right)^2 e^{ik(\bar{\tilde x}\tilde x+\bar{\tilde\xi}\tilde\xi)\tilde u+ik(\bar x'x'+\bar\xi'\xi')u - z(\bar x'\tilde x+\bar\xi'\tilde\xi)+ \bar z(\bar{\tilde x} x'+\bar{\tilde\xi} \xi')}.
$$
The fermionic part needs to be expanded as a Taylor series, extracting only those terms that have exactly one copy of each of $\bar\xi'$, $\xi'$, $\bar{\tilde\xi}$ and $\tilde\xi$, yielding
$$
e^{ik\bar{\tilde\xi}\tilde\xi\tilde u+ik\bar\xi'\xi'u - z\bar\xi'\tilde\xi+ \bar z\bar{\tilde\xi} \xi)}\to -k^2u\tilde u\bar{\tilde\xi}\tilde\xi\bar\xi'\xi'-\bar z z \xi'\tilde\xi\bar{\tilde\xi} \xi.
$$
Upon integration over fermions, this becomes $\bar z z-k^2\tilde u u$, so that
$$
g(X^\dag X)= -\frac{ik\sigma^2\left(\bar z z-k^2\tilde u u\right)X^\dag X}2 \int d\bar x'dx' d\bar{\tilde x}d\tilde x \left(\tilde x-\bar{\tilde x}\right)^2e^{ik\tilde u\bar{\tilde x}\tilde x+iku\bar x'x'- z\bar x'\tilde x+ \bar z\bar{\tilde x} x'}=\frac{k^2\sigma^2u\,X^\dag X}{\bar z z-\tilde u u},
$$
where the $2\times 2$ complex Gaussian integral has been evaluated using (\ref{Gaussboson2}) and the fact that the integral of terms with unequal numbers of barred and non-barred variables vanishes. Then, from (\ref{ggtldtrunc}),
\beq\label{ubranch}
u=\frac{k\sigma^2u}{\bar z z-\tilde u u} \quad\to\quad\mbox{either}\quad u=0\quad \mbox{or}\quad u\tilde u=\bar z z-k\sigma^2.
\eeq
To complete the solution it remains to determine the relation between $u$ and $\tilde u$ that follows from (\ref{gtildegbar}). While we do not have a sufficiently comprehensive theory of the $C$-operator at the moment to treat (\ref{gtildegbar}) systematically, it is easy to guess that the expected circular law results from assuming
\beq\label{utildeubar}
\tilde u =-\bar u.
\eeq
We leave systematic derivation of this formula to future work, and we shall see below that it does not only allow us to rederive the circular law, but also much more sophisticated shapes of spectral edges for more complicated random matrix ensembles in the next section. With (\ref{utildeubar}), $\bar u u=k\sigma^2-\bar z z$, which is only possible inside the circle $|z|^2\le k\sigma^2$, while outside this circle, $u$ must follow the other branch $u=0$, yielding
\beq\label{usol}
u=\begin{cases}
\sqrt{k\sigma^2-\bar z z}&\mbox{if}\quad |z|^2\le k\sigma^2,\\
\hspace{8mm}0&\mbox{otherwise}.
\end{cases}.
\eeq

The solution for $G(z)$ is recovered from (\ref{Gg}) and (\ref{ggtldtrunc}) as 
\beq\label{Geval}
G(z)=
 \displaystyle \int d\bar x dxd\bar\xi d\xi d\bar{\tilde x}d\tilde xd\bar{\tilde\xi }d\tilde \xi \,\bar x\tilde x \,e^{ik(\bar{\tilde x}\tilde x+\bar{\tilde\xi}\tilde\xi)\tilde u+ik(\bar xx+\bar\xi\xi)u - z(\bar x\tilde x+\bar\xi\tilde\xi)+ \bar z(\bar{\tilde x} x+\bar{\tilde\xi} \xi)}.
\eeq
This super-Gaussian integral is structurally identical to the one evaluated immediately above, and one gets the following expression
\beq
G(z)=\frac{\bar z}{\bar z z -\tilde u u}.
\eeq
From (\ref{usol}) and (\ref{ubranch}),
\beq\label{Gsol}
G(z)=\begin{cases}
\bar z/k\sigma^2&\mbox{if}\quad |z|^2\le k\sigma^2,\\
1/z&\mbox{otherwise}.
\end{cases}.
\eeq
Finally, from (\ref{eq:girko}),
\begin{equation}
  \rho(z)=\frac{1}{\pi}\partial_{\bar z}G(z)
  =\frac{1}{\pi k \sigma^2}\,\mathbf{1}\{|z|<\sqrt{k}\sigma\},
  \label{eq:density-inside}
\end{equation}
which is the standard circular law with the radius expressed through the sparsity parameter $k$ and the reference magnitude of the matrix entries $\sigma$.

\subsection{Summary}
\label{sec:susy_spectrum_dictionary}

To recap, starting with the sparse non-Hermitian ensemble defined by \eqref{eq:sparse-ensemble}, we represented its eigenvalue density in terms of the resolvent as in (\ref{eq:girko}), which was then expressed in terms of the Hermitized double-size matrix (\ref{eq:herm-Heps}) and then recast as the Gaussian supervector model 
(\ref{eq:after-average}). This model was solved in the large $N$ limit in terms of the self-consistent saddle point equation (\ref{eq:pf-saddle-no-den}), allowing us to recover the resolvent as (\ref{Gg}). In the limit when the average number of nonzero entries $k$ per row becomes large while the matrix size $N$ is sent to infinity, the Ginibre circular law (\ref{eq:density-inside}) is recovered.

In the next section, we shall apply the methodology developed here to random Jacobians describing equilibria of neuronal networks. This will produce sophisticated shapes of the edges of the spectral distributions, while the density inside will no longer be uniform as it is in (\ref{eq:density-inside}).

We conclude with a brief summary of the computational potential of the theory developed here beyond our concrete practical applications to the shape of the spectral edge at large $k$ that will be explored in the next section. When $k$ is finite, the edge of spectral support is no longer sharp, and the distribution, in principle, extends over the entire complex plane. In practice, the transition towards almost-zero density happens very fast and one obtains a slightly blurred version of the sharp edge observed in the large $k$ limit. (Analysis of the tails of the distribution for related sparse Hermitian matrix problems can be seen in \cite{Rodgers1988density}.)

Various approaches exist for analyzing the finite $k$ case. First, one can try to build expansions in powers of $1/k$, as has been done for sparse Hermitian matrices \cite{Rodgers1988density,Akarapipattana2023Random}, and also for the non-Hermitian case in \cite{Baron2023}. Alternatively, one can attempt to solve the integral saddle point equations numerically, as has been done successfully for the Hermitian case in \cite{Akarapipattana2025Hammerstein,Akarapipattana2025Statistical}. This is normally
preceded by converting the right-hand sides of the superspace equations like (\ref{eq:pf-saddle-no-den}) and (\ref{Gg}) into one-dimensional integrals involving Bessel functions. We explain how to do that in Appendices~\ref{app:proof-rotational-invariant-saddle} and \ref{app:proof-rotational-invariant-G}, but we will not need these equations explicitly in our subsequent exposition and leave their more detailed analysis for future work. This kind of Bessel-kernel saddle point equations has appeared
in the literature on Hermitian matrices \cite{Bray1988,Rodgers1988density, Fyodorov1991, Mirlin_1991,Akarapipattana2023Random,Akarapipattana2025Hammerstein,Akarapipattana2025Statistical, Khetal,onthem,reslarge}, but they have not received the due attention up until now for the non-Hermitian case. (We mention in addition that alternative approaches to numerical solution of related saddle point equations exist based on stochastic sampling in application to multidimensional nonlinear integral equations arising from the `cavity method' approach \cite{Metz2019,sparse2025}.)

One aspect of our construction where further improvement would be welcome is a more explicit understanding of the conjugation relation (\ref{gtildegbar}). For the purposes of our practical derivations in this article, it suffices to assume (\ref{utildeubar}), and no further considerations will be needed. For more general applications,
one would need to understand the spectral properties of the superintegral transform based on the kernel (\ref{eq:C-kernel}). Some studies in this direction can be seen in the literature \cite{Mirlin_1991,transfer1,transfer2}, but the understanding is far from complete. At the same time, for numerical solutions, the space of functions of supervectors will necessarily be truncated to a finite number of dimensions, as in the practical implementations in \cite{Akarapipattana2025Hammerstein,Akarapipattana2025Statistical,localization}. In this case, the said superintegral transform will be represented as a finite-dimensional matrix, and implementing the conjugation (\ref{gtildegbar}) will become straightforward.

\section{Statistical field theory of sparse random Jacobians}
\label{sec:jacobian-SUSY}

Armed with the technology described in the previous section, we now turn to the random Jacobian ensembles that are our primary focus in this paper. In E/I networks of interest, the local Jacobian $J$ around an operating fixed point $\mathbf{x}^*$ encodes stability as in Eq.~\eqref{eq:Jacobian_operating_point}. Recall the structure of the Jacobian matrix $J$ is given by Eq.~\eqref{eq:J_component} 
\begin{equation}
  J_{ij} = \frac{1}{\tau_i}\big(-\delta_{ij} + W_{ij}h_j\big),
  \label{eq:Jacobian-blockdef}
\end{equation}
where $\tau_i>0$ is the synaptic decay timescale of neuron $i$, and $h_j=\sigma'(x^*_j)\in(0,1/4]$ is the gain evaluated at the fixed point, and $W_{ij}$ are the network connectivity. 

We order indices into inhibitory ($I$) and excitatory ($E$) sets with fractions
\beq
f=\frac{|I|}{N},\qquad1-f=\frac{|E|}{N}. 
\eeq
With the shorthands
$\mathcal{T}_E^{-1}=\mathrm{diag}(\tau_{E,i}^{-1})$,
$\mathcal{T}_I^{-1}=\mathrm{diag}(\tau_{I,i}^{-1})$,
$\mathcal{H}_E=\mathrm{diag}(h_{E,i})$, and
$\mathcal{H}_I=\mathrm{diag}(h_{I,i})$, the Jacobian in block form is
\begin{equation}
  J
  =
  \begin{pmatrix}
    -\mathcal{T}_E^{-1} + \mathcal{T}_E^{-1} W_{EE}\mathcal{H}_E
    &
    \mathcal{T}_E^{-1} W_{EI}\mathcal{H}_I
    \\[4pt]
    \mathcal{T}_I^{-1} W_{IE}\mathcal{H}_E
    &
    -\mathcal{T}_I^{-1} + \mathcal{T}_I^{-1} W_{II}\mathcal{H}_I
  \end{pmatrix}.
  \label{eq:J-block-ut}
\end{equation}

For \emph{trained} networks, recall from Sec. \ref{sec: setup_RMT} that the \emph{inhibitory core---excitatory periphery} motif yields
$W_{\bullet E}\approx 0$ (i.e. columns associated with excitatory sources are switched off), so
$W_{EE}\approx 0$ and $W_{IE}\approx 0$. Then $J$ becomes \emph{block‑upper‑triangular}:
\begin{equation}
\label{eq:J-upper-tri}
J \approx
\begin{pmatrix}
-\mathcal{T}_E^{-1} & \mathcal{T}_E^{-1} W_{EI}\mathcal{H}_I\\
0 & -\mathcal{T}_I^{-1} + \mathcal{T}_I^{-1} W_{II}\mathcal{H}_I
\end{pmatrix}.
\end{equation}

\subsection{Spectral decomposition: real bar and inhibitory blob}
Because $J$ of interest, given by \eqref{eq:J-upper-tri}, is block-upper-triangular, its eigenvalues are \emph{exactly} the union of the eigenvalues of the two diagonal blocks. By the visual appearance of the eigenvalue clouds, we refer to the spectrum of $-\mathcal{T}_E^{-1}$ as the \emph{real bar} (the set $\{-1/\tau_{E,i}\}$ on the negative real axis) and to the spectrum of $-\mathcal{T}_I^{-1}+\mathcal{T}_I^{-1}W_{II}\mathcal{H}_I$ as the \emph{inhibitory blob} (a two-dimensional bulk whose rightmost edge sets the distance to marginal stability):
\begin{equation}
  \mathrm{Spec}(J)
  =
  \underbrace{\mathrm{Spec}\left(-\mathcal{T}_E^{-1}\right)}_{\text{real bar}}
  \cup
  \underbrace{\mathrm{Spec}\left(-\mathcal{T}_I^{-1} + \mathcal{T}_I^{-1} W_{II}\mathcal{H}_I\right)}_{\text{inhibitory blob}}.
  \label{eq:spectrum-union-fixed}
\end{equation}
The off‑diagonal block $\mathcal{T}_E^{-1} W_{EI}\mathcal{H}_I$, while having no effect on the eigenvalues, introduces non-normality by mixing the eigenvectors. It allows the inhibitory core to drive the excitatory periphery, which is crucial for how the network's dynamical output is read out. A detailed analysis of the left/right eigenvector structures, and their corresponding roles in shaping network transients and responses to inputs, is  outside the scope of this work focusing on the spectral density.

With this Jacobian ensemble, we can immediately state the {\it spectral decomposition} for this block-upper triangular structure:
\begin{equation}
  \rho_J(z) = \rho_E(z) + \rho_I(z).
  \label{eq:mixture}
\end{equation}

\begin{itemize}
  \item[(\textit{Bar})] The excitatory diagonal block contributes $|E|$ real eigenvalues ($1-f$ fraction of all eigenvalues)
  \begin{equation}
    \lambda_{l} = -\frac{1}{\tau_{E,l}},
    \qquad l\in E,
  \end{equation}
  so the excitatory spectral measure, given the distribution of excitatory timescales $p_{\tau_E}(\tau)$, is
  \begin{equation}
    \rho_E(z) = (1-f)\int d\tau \,p_{\tau_E}(\tau)\delta\left(z+ \frac{1}{\tau}\right).
    \label{eq:bar-density}
  \end{equation}
  \item[(\textit{Blob})] The inhibitory block contributes the remaining $|I|$ eigenvalues ($f$ fraction of all eigenvalues) that extend in the complex plane 
  \begin{equation}
    \rho_I(z) = \mathrm{Spec}\left(-\mathcal{T}_I^{-1} + \mathcal{T}_I^{-1} W_{II}\mathcal{H}_I\right).
    \label{eq:blob-density}
  \end{equation}
\end{itemize}

\noindent
In what follows we analyze the inhibitory blob using the SUSY machinery of section~\ref{sec:intro_susy}.

\subsection{Inhibitory spectral distribution $\rho_I(z)$ via SUSY}
\label{sec:rho_I}

Recall from section~\ref{sec: setup_RMT} that we model $W_{II}$ ($I\,\to\,I$) and $W_{EI}$ ($I\,\to\,E$) as independent sparse random matrices with mean
degrees $k_I$ and $k_E$, respectively, and (possibly different) nonzero-weight distributions.
Because the eigenvalues in \eqref{eq:spectrum-union-fixed} depend only on the diagonal blocks (independent of $W_{EI}$); we therefore focus on the $I\,\to\,I$
block. For $a,b\in I$ we take
\begin{equation}
  \label{eq:W-ensemble}
  \mathbb{P}(W_{ab}) = \left(1-\frac{k_I}{fN}\right)\delta(W_{ab}) + \frac{k_I}{fN}p(W_{ab}),
\end{equation}
where $p(w)$ is the distribution of the nonzero entries. For our concrete application to neuronal networks, this distribution will be a Gaussian with a nonzero (negative) mean. Synaptic timescales are taken i.i.d. from a Beta distribution,
$\tau_{I,i}\sim p_{\tau_I}$ on $I$ and $\tau_{E,i}\sim p_{\tau_E}$ on $E$, and inhibitory gains are i.i.d.
$h_{I,j}\sim p_h$. (Excitatory gains do not affect the eigenvalues in \eqref{eq:spectrum-union-fixed} because
the $E$ block reduces to $-\mathcal{T}_E^{-1}$.)

Following the same steps as in section~\ref{secGauss}, we can represent the resolvent of the inhibitory Jacobian $G(z)\equiv-\mathbb{E}\Tr(J_I-zI)^{-1}/N=-\mathbb{E}(J_I-zI)^{-1}_{11}$ as the following super-Gaussian integral:
$$
  G(z)=   \mathbb{E}\!\!\int [dX]\,[d\tilde X]\, \bar{x}_{1}\tilde{x}_{1} \exp\bigg\{\sum_{a,b}W_{ab}\left(X^{\dag}_{a}\tilde{X}_{b}-\tilde{X}^{\dag}_{b}X_{a}\right)h_{b}/\tau_{I, a} -\sum_{a}\left[(z+1/\tau_{I, a})X^{\dag}_{a}\tilde{X}_{a}-(\bar{z}+1/\tau_{I, a})\tilde{X}^{\dag}_{a}X_{a}\right] \bigg\}.
$$
Unlike the simplest sparse non-Hermitian ensemble that we have treated as a warm-up, here, the averaging over $W$ cannot be performed immediately,
as it is not disentangled from the averaging over the temporal heterogeneity $\tau_I$ and the synaptic gains $h$.
A simple supervector rescaling $X_{a}\rightarrow X_{a}\tau_{I,a}$,  and $\tilde{X}_{a}\rightarrow \tilde{X}_{a}/h_{a}$, however, achieves the necessary decoupling:
\beq
    G(z)=\mathbb{E} \!\!\int  [dX]\,[d\tilde X]\,  \frac{\tau_{I,1}}{h_1}\bar{x}_{1}\tilde{x}_{1} \exp\bigg\{\!\!\sum_{a,b}W_{ab}\left(X^{\dag}_{a}\tilde{X}_{b}-\tilde{X}^{\dag}_{b}X_{a}\right) -\sum_{a}\left[(z\tau_{I,a}\!+1\!)X^{\dag}_{a}\tilde{X}_{a}\!-\!(\bar{z}\tau_{I,a}\!+\!1)\tilde{X}^{\dag}_{a}X_{a}\right]/h_{a} \bigg\}. \label{eq:resolvent_trained_Jacobian_2}
\eeq
(Note that a uniform rescaling of a supervector does not generate any new factors in the integration measure, since such factors cancel out between the bosonic and fermionic components.)
Starting with the above representation, one can straightforwardly proceed with averaging over $W$, $\tau_I$ and $h$, since the integrand is completely factorized over sites.
Thereafter, we can apply the Fyodorov-Mirlin decoupling (\ref{eq:FM-identity}) to the terms arising from averaging over $W$ and arrive at a saddle point equation as in the previous section. A convenient way to think about the structure is as follows: in the derivations of the previous section, one must replace $z$ with $(z\tau_{I}+1)/h$. But since this new expression involves random variables $\tau_I$ and $h$, the resulting expressions must be averaged over those distributions.
Proceeding in this manner, we arrive, instead of (\ref{eq:pf-saddle-no-den}), at the following saddle point equation
\begin{equation}
  \label{eq:pf-saddle-tauh}
g(X)= i k_I\, \mathbb{E}_{\tau_I,h}\int dX' d\tilde X \int_{\mathbb{R}} dw\,p(w) \left[1-e^{w(X^\dag \tilde X-\tilde X^\dag X)} \right]\, e^{i\tilde g(\tilde X)+ig(X') - (z\tau_{I}+1) X'^\dag \tilde X /h+ (\bar z\tau_{I}+1) \tilde X^\dag X'/h},
\end{equation}
where $C$ has been substituted from (\ref{eq:C-kernel}), and assuming $g(X)=g(X^\dag X)$ and $\tilde g(X)=\tilde g(X^\dag X)$. From the solution of this equation, one recovers the resolvent as
\begin{equation}\label{restauh}
    G(z)=\mathbb{E}_{\tau_I,h}\int dX  d\tilde X \,\frac{\tau_{I}}{h}\,\bar x\tilde x\,
e^{i\tilde g(\tilde X)+ig(X) - (z\tau_{I}+1) X^\dag \tilde X /h+ (\bar z\tau_{I}+1) \tilde X^\dag X/h}.
\end{equation}
Equations (\ref{eq:pf-saddle-tauh}) and (\ref{restauh}) provide a systematic framework for studying spectral densities of sparse Jacobians. They can be reduced to one-dimensional integral equations and explored by means of $1/k$ expansions and numerical solution as outlined in the previous section, but we will not pursue those derivations here.

Our main focus here is on the {\it spectral edge} which plays a crucial role for the near-critical stability, which is in turn essential for successful operation
of the neuronal network. The edge becomes sharp in the large $k_I$ limit, while at finite $k_I$ it is somewhat smeared. The actual smearing for the values of $k_I$ of order 10 (which is what concerns us here) is, however, small, and the shape of the distributions is accurately captured by the leading-order formulas at large $k_I$, as we shall see explicitly in the numerical examples below.

At large $k_I$, the integrals on the right-hand side of equation (\ref{eq:pf-saddle-tauh}) acquire a saddle point structure, as described in section~\ref{subsec:saddle_to_spectrum}, and at leading order, $g$ and $\tilde g$ can be self-consistently approximated as
\beq
g(X)=k_I\,u\,X^\dag X,\qquad \tilde g(X)=k_I\,\tilde u\,X^\dag X.
\eeq
With this ansatz, the integrals on the right-hand side of (\ref{eq:pf-saddle-tauh}) become super-Gaussian, and are evaluated explicitly, resulting in an algebraic equation
\beq
u=u\,k_I\sigma^2\,\mathbb{E}_{\tau_I,h}\left[\frac{1}{|z\tau_{I}+1|^2/h^2-\tilde u u}\right]
\eeq
analogous to (\ref{ubranch}), where again $\sigma^2 \equiv\int dw\, p(w)\,w^2$. There are two solutions: the trivial solution $u=0$ and the other solutions respecting 
$k_I\sigma^2\,\mathbb{E}_{\tau_I,h}\left(|z\tau_{I}+1|^2/h^2-\tilde u u\right)^{-1}=1$, where one must keep in mind that $\tilde u=-\bar u$ by (\ref{utildeubar}). The spectral edge is located precisely where the nontrivial branch touches the trivial one with $u=0$, which means that
\begin{equation}
\label{eq:inh-edge-condition}
k_I \sigma^2 \,\,\mathbb{E}_{h}[h^2]\,\,\,\mathbb{E}_{\tau_I}\left[\frac{1}{|z\tau_I+1|^2}\right] = 1.
\end{equation}
This {\it spectral edge condition} then determines the curve that bounds the spectral support as a {\it level set} of the function $\mathbb{E}_{\tau_I}|z\tau_I+1|^{-2}$ in the complex plane, obtained as a simple average over the ensemble of heterogeneous inhibitory timescales. In what follows, we will test the operation of this spectral edge condition in a few relevant examples.

\subsection{Homogeneous $\tau_I$: the circular inhibitory spectral edge}
\label{sec:homo-tau}

\begin{figure}
    \centering
    \includegraphics[width=0.85\linewidth]{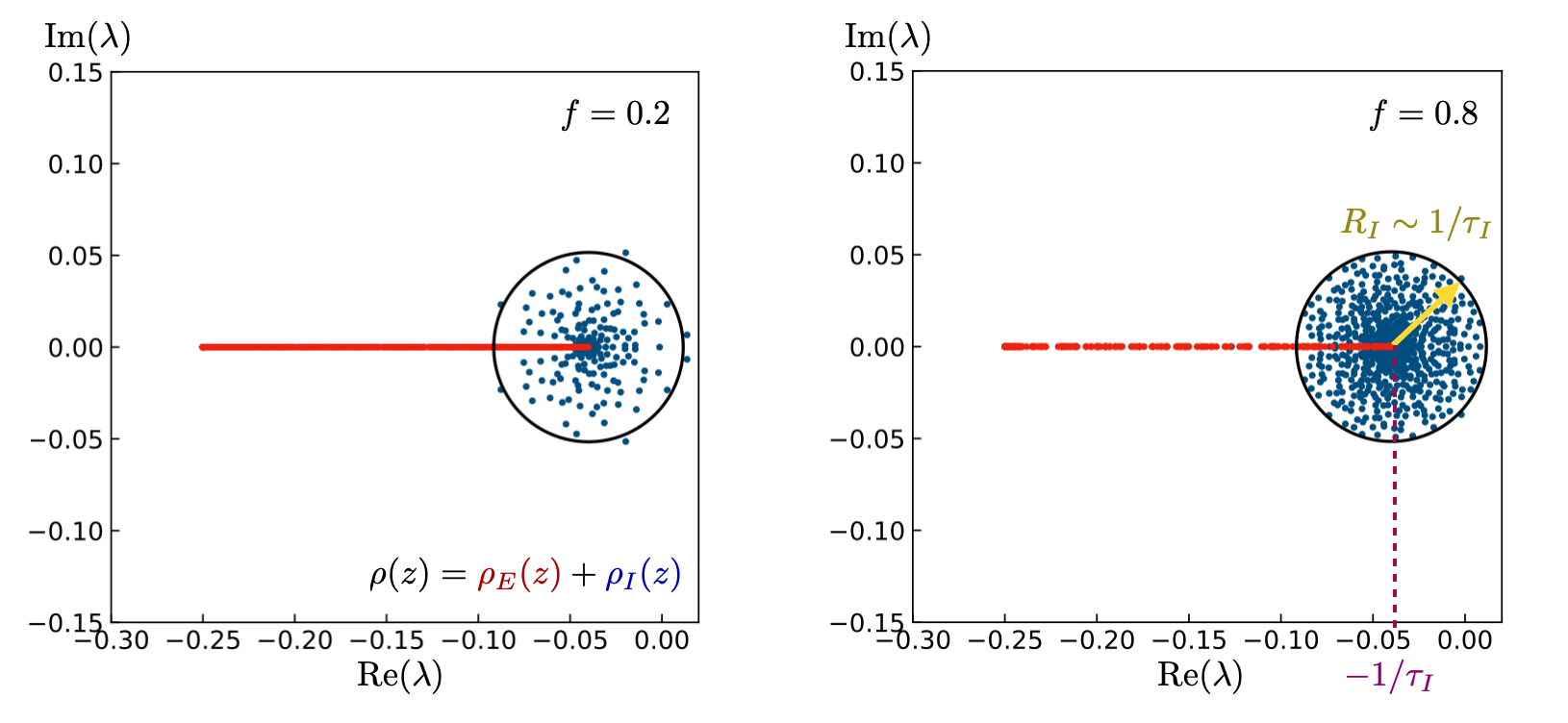}
    \caption{
\textbf{The spectrum for homogeneous inhibitory timescale and robust marginality from slow inhibition.}
The Jacobian spectrum $\rho(z)$ decomposes into the real-valued excitatory bar $\rho_E(z)=\{1/\tau_{E,l} \ | \ l\in E\}$ (red dots) and the complex-valued inhibitory blob $\rho_I(z)$ (blue dots): $\rho(z) = \rho_E(z) + \rho_I(z)$.
\textbf{(Left)} Spectrum with a biologically realistic (cortical area) inhibitory fraction ($f=0.2$). \textbf{(Right)} Spectrum with a large inhibitory fraction ($f=0.8$). The spectral edge of the inhibitory blob is independent of $f$. In both plots, the analytic theory for a homogeneous timescale ($\tau_I = 25$) correctly predicts the blob as a disk centered at $C_I = -1/\tau_I = -0.04$ (dashed magenta line) with a radius $R_I = \tau_I^{-1}\sqrt{k_I\sigma^2\mathbb{E}_h[h^2]}$ (yellow arrow), as derived in Eq.~\eqref{eq:radius-homogeneous}.
Eigenvalues (dots) are from exact diagonalization of a single realization with $N=1000$.
Sparse connectivity parameters $k_I$ and weight statistics ($\mu, \sigma^2$) were extracted from trained sample \texttt{000815} ($N_{\text{train}}=200$) of \cite{Rungratsameetaweemana2025}. }
\label{fig:homogeneous_tau_I}
\end{figure}

A particularly transparent case for the spectral edge condition is constant $\tau_I$. Then \eqref{eq:inh-edge-condition}
becomes
\begin{equation}
  \left|z + \frac{1}{\tau_I}\right| = \frac{1}{\tau_I}\sqrt{k_I\sigma^2\mathbb{E}_h[h^2]}.
\end{equation}
Thus, the inhibitory bulk spectrum is contained within a disk centered at $-1/\tau_I$ with radius
\begin{equation}
  R_I = \frac{\sqrt{k_I\sigma^2\mathbb{E}_h[h^2]}}{\tau_I}.
  \label{eq:radius-homogeneous}
\end{equation}
The rightmost point of the inhibitory blob on the real axis is
\begin{equation}
  \Re z_{\max}
  = \frac{\sqrt{k_I\sigma^2\mathbb{E}_h[h^2]} - 1}{\tau_I}.
  \label{eq:rightmost}
\end{equation}
A \emph{near-marginality criterion} for this uniform inhibitory timescale is given by $\Re z_{\max} \approx 0^-$, which requires tuning
\begin{equation}
\sqrt{k_I\sigma^2\mathbb{E}_h[h^2]}\approx 1^-.
\label{eq:near-marginal-criterion}
\end{equation}
These spectra are shown in Fig.~\ref{fig:homogeneous_tau_I}. Note that the density is not uniform inside the spectral edge circle. This picture is reminiscent of the very basic random matrix ensembles used at the very beginning of discussions on stability criteria for systems with random couplings \cite{May1972}.

\begin{figure}[h!]
    \centering
    \includegraphics[width=0.85\linewidth]{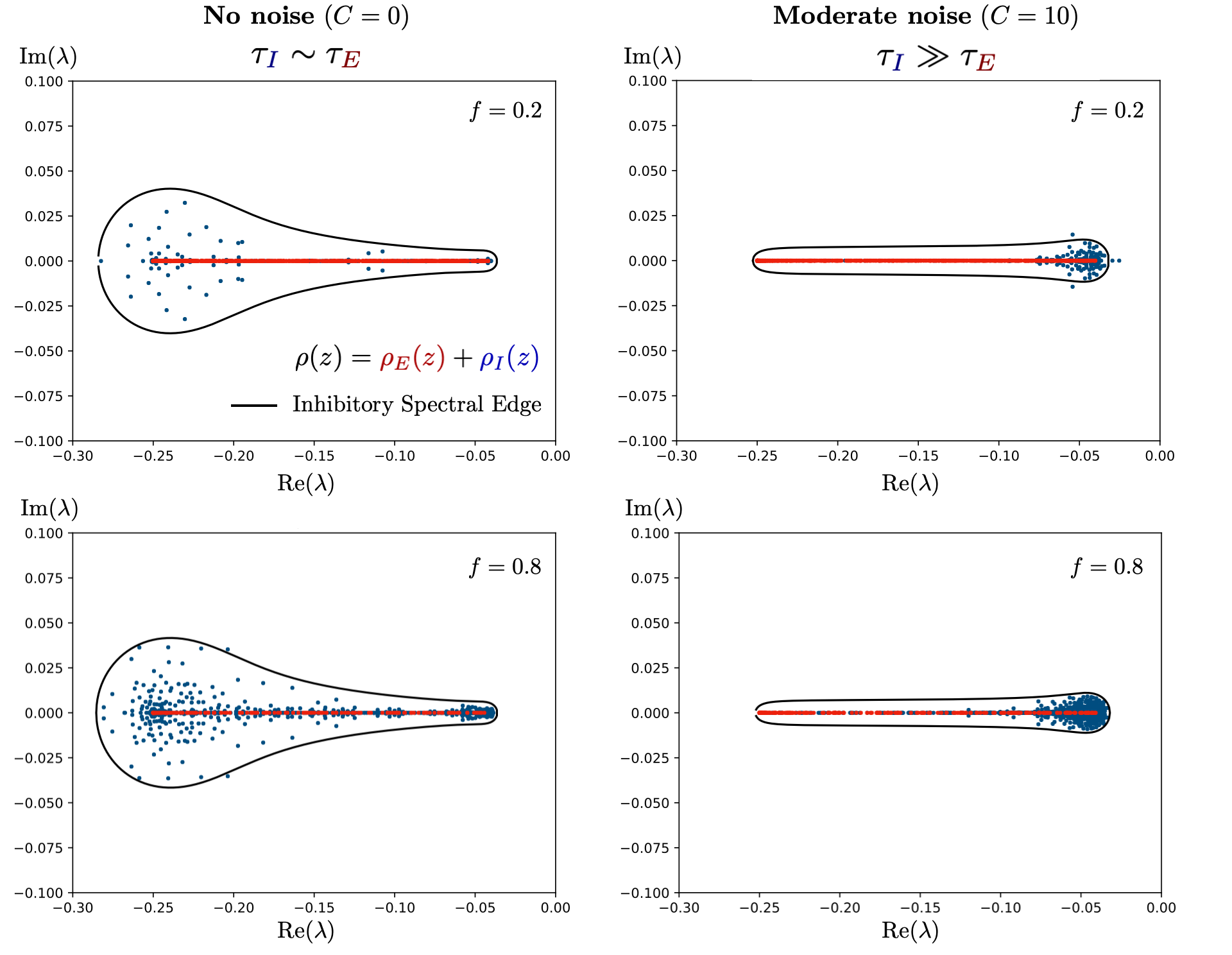}
    \caption{
\textbf{Validation of the spectral theory of the sparse Jacobian ensemble against trained networks statistics.}
Comparison of numerical eigenvalues (dots, exact diagonalization with $N=1000$) against the theoretical \emph{inhibitory spectral edge} (black solid line) predicted by Eq.~\eqref{eq:inh-edge-condition}.
\textbf{(Left Column)} Network trained without noise ($C=0$, sample \texttt{114930} of \cite{Rungratsameetaweemana2025}). The fitted timescale distributions lack separation ($\tau_I \sim \tau_E$), and the inhibitory blob is far from the instability boundary.
\textbf{(Right Column)} Network trained with moderate noise ($C=10$, sample \texttt{000815} of \cite{Rungratsameetaweemana2025}). The emergence of slow inhibitory timescales ($\tau_I \gg \tau_E$) compresses the spectrum and pushes the inhibitory blob toward the origin, achieving near-marginality.
\textbf{(Rows)} The theory accurately predicts the envelope for both the biologically relevant inhibitory fraction ($f=0.2$, top) and a synthetic case ($f=0.8$, bottom), showing that the spectral edge depends on sparsity $k_I$ rather than the inhibitory fraction $f$.
Here the Jacobian matrices were generated using statistics (sparsity $k_I$, weight moments, gain moments, and timescale fits) extracted from the specified trained samples ($N_{\text{train}}=200$).}
    \label{fig: edge_validation}
\end{figure}

\begin{figure}[h!]
    \centering
    \includegraphics[width=0.5\linewidth]{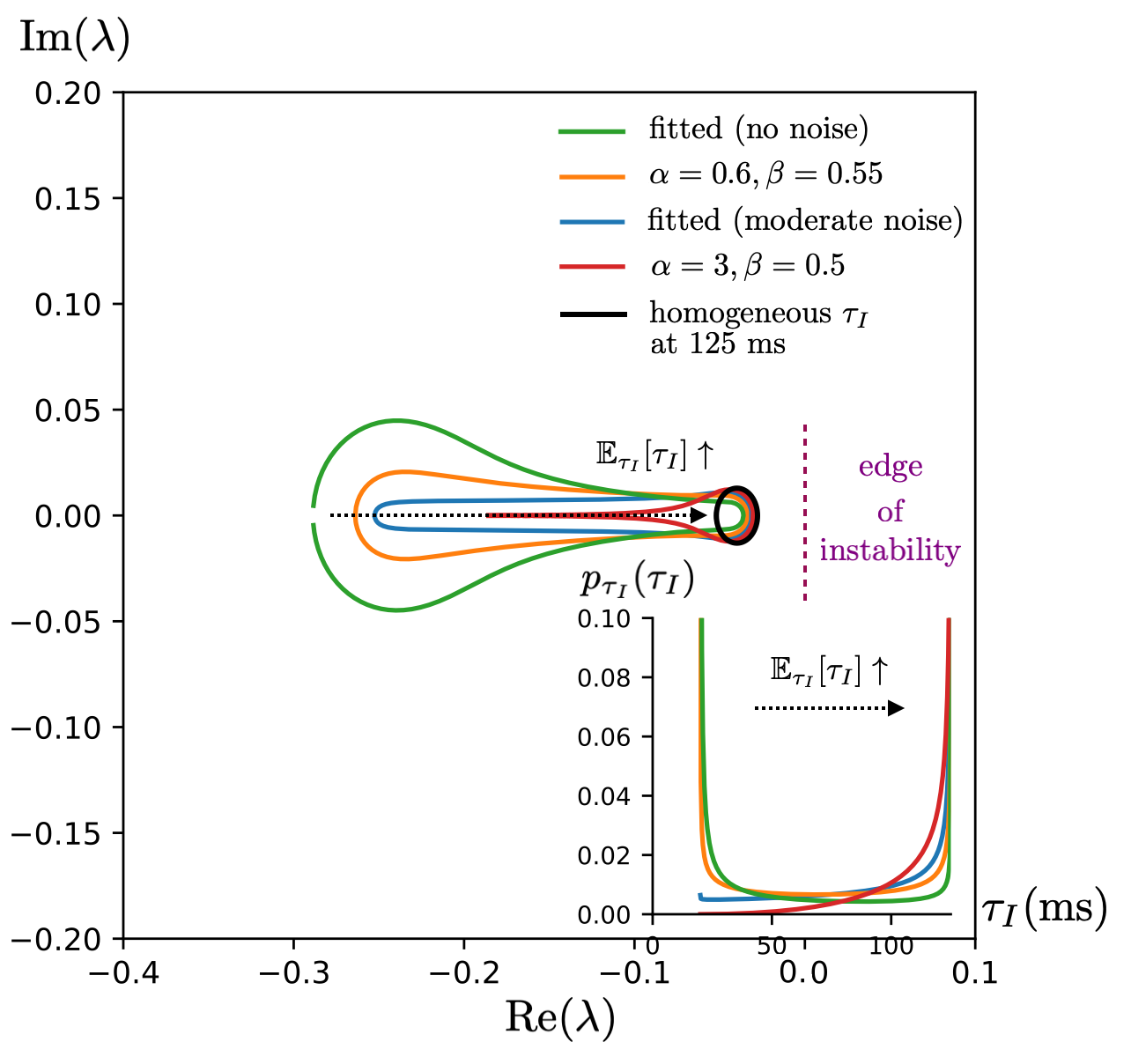}
    \caption{
\textbf{Slow inhibitory timescales push the inhibitory spectral edge toward marginality.} 
    This figure illustrates how the shape of the inhibitory timescale distribution, $p_{\tau_I}(\tau_I)$ (inset), influences the shape of the inhibitory spectral edge (main plot), as predicted by Eq.~\eqref{eq:inh-edge-condition}. The inset shows several Beta distributions for $\tau_I$ on the biological range $[25, 125]$ ms, while the main plot shows the corresponding theoretical spectral boundaries. The sparsity $k_I$, weight variance $\sigma^2$, and gain statistics $\mathbb{E}[h^2]$ are fixed to values fitted from the moderate-noise trained network ($C=10$, sample \texttt{000815} of \cite{Rungratsameetaweemana2025}). Only $p_{\tau_I}$ is varied across curves: \textbf{(i) Green (Fitted,  $C=0$):} Timescale distribution fitted from sample \texttt{114930} of \cite{Rungratsameetaweemana2025}. Heavily skewed toward fast timescales (low $\mathbb{E}[\tau_I]$), resulting in a spectral edge far from the instability boundary at $\mathrm{Re}(\lambda)=0$; \textbf{(ii) Blue (Fitted, $C=10$):} Timescale distribution fitted from sample \texttt{000815} of \cite{Rungratsameetaweemana2025}. Training with moderate noise shifts the distribution toward slower timescales (higher $\mathbb{E}[\tau_I]$), moving the spectral edge closer to marginality; \textbf{(iii) \& (iv) Orange \& Red (Synthetic):} U-shaped (orange) and heavily skewed slow (red) distributions demonstrate that increasing the density of slow timescales further pushes the spectral boundary toward marginality; \textbf{(v) Black (Homogeneous $\tau_I$):} Limiting case of a Dirac delta at $\tau_I = 125$ ms, recovering the circular edge from Sec.~\ref{sec:homo-tau}. The arrows (inset and main) summarize the key trends: as the mean inhibitory timescale $\mathbb{E}_{\tau_I}[\tau_I]$ increases, the entire inhibitory spectrum is systematically pushed toward the edge-of-stability, while the heterogeneity of these timescales simultaneously enables a broad relaxation spectrum.
    }
    \label{fig: hetero_tau_edge_shape}
\end{figure}

\subsection{Heterogeneous $\tau_I$: properties of the rightmost spectral edge}
\label{sec:hetero-tau}

In general, \eqref{eq:inh-edge-condition} defines a complicated shape in the $z$-plane that one can straightforwardly reconstruct numerically, as in the examples below. Some general properties of this shape, and in particular its closest approach to the imaginary axis that controls near-critical stability, can be stated as follows:

\paragraph*{Rightmost edge lies on the real axis:} Consider $z=x+iy$ and write \eqref{eq:inh-edge-condition} as
\begin{equation}
  F(x,y)\equiv \mathbb{E}_{\tau_I}\left[\frac{1}{(x\tau_I+1)^2+(y\tau_I)^2 }\right] =  \left( k_I \sigma^2\,\mathbb{E}_{h}[h^2] \right)^{-1}.
  \label{eq:rightmost-hetero}
\end{equation}
The edge is a level set of $F(x,y)$. Importantly, $F$ is a strictly decreasing function of $|y|$ since $1/(a^2+y^2)\le 1/a^2$ for any $a$, and the equality is only attained when $y=0$. The level set cannot extend to $x=+\infty$ since $F$ decays there --- therefore, there must exist the maximal value of $\Re z$ that we call $x_0$ attained by the level set. If that happens at the point $x_0+iy_0$, the smooth level set curve must extend away from that point into the quadrants $(x<x_0,y<y_0)$ and $(x<x_0, y>y_0)$. But no two points with the same $x$ and two different values of $y$ can belong to the same level set by the above property of monotonic decrease. The only way to avoid contradiction is $y_0=0$, since $(x,y)$ and $(x,-y)$ do belong to the same level set. Therefore $y_0=0$, and the maximum of $\Re z$ is attained on the real axis.

\paragraph*{Lower bound on the rightmost inhibitory edge:}
Since the rightmost edge occurs on the real axis, so we set $z=x\in\mathbb{R}$ and write \eqref{eq:rightmost-hetero} as
\[
\mathbb{E}_{\tau_I}\left[\frac{1}{(x\tau_I+1)^2}\right]=\kappa^{-1}\equiv \left( k_I \sigma^2\,\mathbb{E}_{h}[h^2] \right)^{-1}.
\]
Since $t\mapsto 1/t$ is convex on $(0,\infty)$, Jensen’s inequality $\mathbb{E}\left[{1}/{X}\right]\ge{1}/{\mathbb{E}[X]}$ gives
\[
\mathbb{E}[(x\tau_I+1)^2]\ge\kappa.
\]
Expanding the left-hand side and using $\mathbb{E}[\tau_I]=\mu_{\tau_I}$,
$\mathbb{E}[\tau_I^2]=\mu_{\tau_I}^2+\sigma_{\tau_I}^2$ yields
\[
(\mu_{\tau_I}^2+\sigma_{\tau_I}^2)x^2 + 2\mu_{\tau_I}x + 1 \ge \kappa,
\]
Since $\mu_{\tau_I}^2+\sigma_{\tau_I}^2>0$, the solution lies in $(-\infty,x_-]\cup[x_+,\infty)$ with
\[
x_\pm = \frac{-\mu_{\tau_I} \pm \sqrt{\mu_{\tau_I}^2 - (\mu_{\tau_I}^2+\sigma_{\tau_I}^2)(1-\kappa)}}{\mu_{\tau_I}^2+\sigma_{\tau_I}^2}.
\]
The rightmost edge $x_\star$ must lie in the right interval, hence $x_\star\ge x_+$ whenever the
discriminant is non-negative, i.e., when $(1-\kappa) \le \mu_{\tau_I}^2/(\mu_{\tau_I}^2+\sigma_{\tau_I}^2)$. For $\sigma_{\tau_I}^2=0$ the Jensen's inequality
becomes an equality (Jensen's inequality is tight on a Dirac delta distribution), giving
$x_\star=(-\mu_{\tau_I}+\mu_{\tau_I}\sqrt{\kappa})/\mu_{\tau_I}^2=(\sqrt{\kappa}-1)/\mu_{\tau_I}$, in agreement with the homogeneous case in Sec.~\ref{sec:homo-tau}.

One can rewrite the lower bound as 
\begin{equation}\label{eq: lower_bound_rightmostedge}
x_\star \ge x_+= \frac{1}{\mu_{\tau_I}} \left( \frac{1}{1+\sigma^2_{\tau_I}/\mu^2_{\tau_I}} \right) \left(- 1 + \sqrt{1-(1+\sigma^2_{\tau_I}/\mu^2_{\tau_I})(1-\kappa)}\right).
\end{equation}
The prefactor $1/\mu_{\tau_I}$ sets the overall timescale, while the synaptic heterogeneity ($\sigma_{\tau_I}$) plays a more nuanced role. In Fig.~\ref{fig: hetero_tau_edge_shape}, we present a collection of spectral edge shapes for the $\tau_I$-distributions extracted from trained networks, and the Beta-distributions
that approximate them, as well as some synthetic $\tau_I$ distributions. In this context, increasing $\sigma_{\tau_I}$ stretches the bulk spectrum along the real axis: fast inhibitory timescales (small $\tau_I$) contribute modes with strongly negative real parts, while slow timescales (large $\tau_I$) pull the rightmost edge $x_\star$ closer to the imaginary axis. The resulting inhibitory blob thus realizes a \emph{broad relaxation spectrum} in which deeply stable modes coexist with near-marginal ones.

\section{Discussion and outlook}\label{sec:outlook}

We have developed a sparse non-Hermitian random matrix theory to analyze the Jacobian spectra governing the relaxation dynamics of recurrent neuronal networks that perform robust working memory computation via discrete-attractor hopping \cite{Rungratsameetaweemana2025}, as described in Sec.~\ref{sec: working memory}. These trained networks exhibit an \emph{inhibitory core---excitatory periphery} architecture characterized by broad heterogeneity of synaptic timescales and sparse connectivity. By formalizing this structure, the spectrum decomposes into a trivial excitatory bar and an inhibitory blob as in Eq.~\eqref{eq:spectrum-union-fixed}, and we have derived an analytic condition for the inhibitory spectral edge, given by Eq.~\eqref{eq:inh-edge-condition}, that depends on inhibitory timescale distributions, as well as weight and gain functions statistics. 
Our studies of the statistics of these neuronal network parameters, based on the trained networks developed in \cite{Rungratsameetaweemana2025},
indicate that the relevant distributions are well-fitted by very simple curves: a Gaussian distribution for the nonzero inhibitory weights, and Beta distributions
for the synaptic timescales and gains. The analytic random matrix description we have developed here allows us to convert these statistical specifications into predictions about the spectral edge, and in particular, about whether the edge-of-stability property is respected.
Numerical results further suggest that slow inhibition pushes the spectrum to cluster near the rightmost edge (Fig.~\ref{fig: edge_validation}, right column). Such clustering implies an increased density of near marginally stable modes available for state transitions, a favorable feature that deserves further investigation.

Regarding the roles of eigenvectors, numerical results in Ref.~\cite{Rungratsameetaweemana2025} indicate that the left eigenfunctions associated with near-marginal modes, which govern the sensitivity to perturbations, are relatively delocalized. This might suggest a robustness mechanism akin to a ``lock-and-key'' system: while the modes are soft, destabilizing the attractor requires a precise, distributed perturbation across many neurons to drive the system into the basin of the next attractor. Validating this hypothesis requires an analytic treatment of the Inverse Participation Ratio (IPR) of the left eigenfunctions in this random Jacobian ensemble, which involves computing `four-point functions' in the corresponding supervector model, a further step from the `two-point functions' (insertions of $\bar x\tilde x$) used in our analysis of the spectral density. Furthermore, while the spectral role of the inhibitory core is clear, the functional role of excitatory units is not captured by their relaxation spectrum alone, as they form a trivial real bar. Perhaps, the functional contribution of excitatory units might arise from the \emph{non-normality} inherent in the off-diagonal block $W_{EI}$ in Eq.~\eqref{eq:J-upper-tri}. Such non-normality can lead to transient amplification, where the non-orthogonality of the eigenvectors allows specific perturbations to facilitate escape from the basin \cite{nowak_nonorthogonal, nowak_transientamp}. Investigating these effects, alongside determining the memory capacity supported by the inhibitory core---excitatory periphery motif (analogous to counting the number of stable fixed points, or memory patterns, in Hopfield networks), is essential for a complete understanding of working memory computation in this discrete attractor network ensemble.

From the perspective of random matrix theory, our contribution is, most broadly, in showing a pathway to systematically employing supersymmetry-based techniques
for analyzing sparse non-Hermitian random matrix ensembles. This approach recasts the answers to various questions, in the limit where the matrices become large,
in terms of self-consistent integral equations that can be later treated by asymptotic or numerical methods. Even for dense matrices, where other methods \cite{Rajan2006,lindbladian,spindle} are available
(primarily, based on the method of moments, including its modern reincarnations in free probability theory), our methodology provides an elegant and economical alternative. One demonstration is given in Appendix~\ref{app:RAM-like_model} where we rederive with minimal effort the eigenvalue density of the Rajan-Abbott ensemble \cite{Rajan2006}. When the matrices become sparse, applications of the method of moments meet substantial difficulties, while the technology presented here
continues to work, and provides access to a variety of questions. As the number of nonzero entries in sparse matrices grows, the spectrum approaches in shape to that of
dense matrices. Our representation is well-adapted to capturing this transition, and we have used this feature to obtain a simple and explicit theory of the phenomenologically important spectral edge
(which becomes sharp as the number of nonzero entries grows). At the same time, more refined questions directly involving sparsity may be systematically addressed in our formalism, and we leave that to future work.

Finally, from the perspective of dynamics of complex systems, this study reinforces the notion that incorporating realistic constraints \cite{Nemenman2025}, such as timescale heterogeneity and structural motifs, can lead to new functional mechanisms that control near-critical properties. The idea that statistical properties of random matrices
get translated into stability conditions for dynamical systems with couplings given by these random matrices goes back to the foundational work of May \cite{May1972} on large ecosystems. Sparsity of couplings is one criterion for stability that emerges from this perspective as it compresses the overall extent of the spectrum.
Here, such sparsity of couplings indeed emerges naturally from training neuronal networks treated in \cite{Rungratsameetaweemana2025}.
(Another recent discussion of the role of sparsity in neuronal networks can be seen in \cite{sparsity}, while more refined studies of real-time dynamics of random non-Hermitian systems can be seen in \cite{transient1,transient2}.)
In the context of neural networks performing robust working memory computation, the functional requirement is more stringent than in applications to ecology: stability alone is insufficient, as it can cause the incoming input to fade too quickly and the operating fixed point to become too stable to be driven by the subsequent input, as discussed in Sec.~\ref{sec: working memory}. What one needs instead is near-critical stability of typical operating fixed points. Mathematically, this functional requirement tightens a broad stability statement in ecology (an inequality) to a precise tuning requirement (an approximate equality) where the spectral edge hovers near the imaginary axis. Our results here analytically derive this condition, explicitly relating the edge-of-stability to the statistical properties of synaptic weights, timescales, and gains.\newpage

\begin{acknowledgments}
We thank E. Gudowska-Nowak,  M. Nowak, R. Kim and N. Rungratsameetaweemana for valuable discussions.  T.C. and W.H. additionally thank E. Gudowska-Nowak and M. Nowak for their hospitality during our visit to the Mark Kac Complex Systems Research Center at Jagiellonian University. T.C. and O.E. acknowledge funding support from the NSRF via PMU-B, grant number B13F670063. O.E. also acknowledges support from the  C2F program at Chulalongkorn University and by NSRF via grant number B41G680029,
and additionally from a Priority Research
Area DigiWorld grant under the Strategic Programme
Excellence Initiative at the Jagiellonian University
(Krak\'ow, Poland). W.H. acknowledges support 
from the National Science
and Technology Development Agency under the JSTP
scholarship, grant number SCA-CO-2565-16992-TH.
\end{acknowledgments}

\appendix

\section{Empirical Statistics of Trained Network Characteristics}
\label{app:network_stats}

To validate the assumptions underlying the sparse non-Hermitian ensemble defined in Sec.~\ref{sec: setup_RMT}, we extracted the statistical properties of the trained recurrent weights $W$ obtained from the public repository accompanying Ref.~\cite{Rungratsameetaweemana2025} (\url{https://osf.io/dqy3g/}). The statistics summarized below (averaged over 50 trained network realizations) confirm the emergence of the inhibitory core---excitatory periphery motif.

Table~\ref{tab:exc_pruning} demonstrates the extreme pruning of excitatory connections. Even under high-noise training ($C=50$), the number of non-zero excitatory weights is negligible compared to the total number of possible excitatory synapses. Consequently, the approximation $W_{\bullet E} \approx 0$ used in the RMT formulation is reasonable.

Tables~\ref{tab:inh_weights} and \ref{tab:inh_sparsity} characterize the inhibitory block. The inhibitory weights follow a stable distribution with negative means (Table~\ref{tab:inh_weights}), and the connectivity remains sparse (Table~\ref{tab:inh_sparsity}), supporting the modeling of the inhibitory block as a sparse random matrix with finite mean degree.

\begin{table}[h]
    \centering
    \setlength{\tabcolsep}{8pt} 
    \begin{tabular}{ccccccc}
        \hline \hline
        Statistic & No noise & $C=1$ & $C=5$ & $C=10$ & $C=20$ & $C=50$ \\ 
        \hline
        Mean & 8.82 & 9.26 & 6.56 & 6.62 & 31.80 & 72.18 \\
        Std. Dev. & 4.31 & 6.01 & 4.69 & 5.10 & 23.77 & 31.31 \\
        \hline \hline
    \end{tabular}
    \caption{\textbf{Count of non-zero excitatory connections.} The table shows the average number of surviving excitatory weights ($W_{ij} > 0$) across different noise channels (noise levels) $C$. Note that for a network of size $N=200$ with $80\%$ excitatory neurons, the total number of possible excitatory connections is $N \times N_E \approx 32,000$. Thus, even at $C=50$, the excitatory connectivity is effectively negligible ($\ll 1\%$ density), validating the pruned excitatory periphery assumption.}
    \label{tab:exc_pruning}
\end{table}

\begin{table}[h]
    \centering
    \setlength{\tabcolsep}{8pt}
    \begin{tabular}{ccccccc}
        \hline \hline
        Statistic & No noise & $C=1$ & $C=5$ & $C=10$ & $C=20$ & $C=50$ \\ 
        \hline
        Mean & -7.294 & -6.739 & -5.500 & -4.866 & -4.535 & -5.377 \\
        Std. Dev. & 2.898 & 2.671 & 2.146 & 1.840 & 1.688 & 1.987 \\
        \hline \hline
    \end{tabular}
    \caption{\textbf{Statistics of non-zero inhibitory weights.} The mean and standard deviation of the non-zero weights in the inhibitory columns ($W_{\bullet I}$). The consistent negative mean and bounded variance across noise conditions support the Gaussian modeling of non-zero inhibitory weights.}
    \label{tab:inh_weights}
\end{table}

\begin{table}[h]
    \centering
    \setlength{\tabcolsep}{8pt}
    \begin{tabular}{ccccccc}
        \hline \hline
        Connection Type & No noise & $C=1$ & $C=5$ & $C=10$ & $C=20$ & $C=50$ \\ 
        \hline
        $I \to I$ ($k_I/N_{I}$) & 0.123 & 0.135 & 0.131 & 0.137 & 0.103 & 0.107 \\
        $I \to E$ ($k_E/N_{I}$) & 0.199 & 0.190 & 0.195 & 0.192 & 0.168 & 0.202 \\
        \hline \hline
    \end{tabular}
    \caption{\textbf{Connectivity density of the inhibitory core.} The table reports the connection probability, defined as the mean degree $k$ normalized by the size of the inhibitory population ($N_I$). $k_I$ represents inhibitory-to-inhibitory connections, and $k_E$ represents inhibitory-to-excitatory connections. The relatively low values ($\sim 0.1 - 0.2$) confirm that the inhibitory block remains sparse.}
    \label{tab:inh_sparsity}
\end{table}

\section{Details for the analytic treatment of the general non-Gaussian sparse ensemble}
\label{app:warmup-general}

\subsection{SUSY integral and disorder average}\label{app:susy}

We begin by taking the expectation of Eq.~\eqref{eq:resolvent-elem} over the ensemble of $M$. Since the integral is over variables that are independent of the randomness, the expectation can be brought inside the integral as
\begin{equation}
-\mathbb{E}(M-zI)^{-1}_{11} = \int \big(\prod_a dX_ad\tilde X_a\big)\bar x_1\tilde x_1  \mathbb{E}_M\left( \exp\left[ \ 
\underbrace{\sum_{a,b} \Big( X_a^\dag (M - zI)_{ab} \tilde X_b - \tilde X_a^\dag (M^{T} - \bar z I)_{ab} X_b \Big)
}_{\mathrm{Exponent}}
\ \right] \right),
\end{equation}
where $M$ is a real matrix, so $M^\dagger = M^T$. 
Separate the terms in the exponent that depend on $M$ from the terms that do not as
\begin{align}
\text{Exponent} &= \sum_{a,b} \left( X_a^\dag (M_{ab} - z\delta_{ab}) \tilde X_b - \tilde X_a^\dag (M_{ba} - \bar z\delta_{ab}) X_b \right)  \\
&= \underbrace{ \sum_{a,b} \left( M_{ab} X_a^\dag \tilde X_b - M_{ba} \tilde X_a^\dag X_b \right) }_{M\text{-dependent exponent} } \underbrace{ - \sum_{a,b} \left( z X_a^\dag \delta_{ab} \tilde X_b - \bar z \tilde X_a^\dag \delta_{ab} X_b \right) }_{M\text{-independent exponent} }. 
\end{align}
The $M$-independent exponent can be pulled out of the expectation $\mathbb{E}_M[\cdot]$, and simplifies to
\begin{equation}
M\text{-independent exponent} = - \sum_{a=1}^N \left( z X_a^\dag \tilde X_a - \bar z \tilde X_a^\dag X_a \right),
\end{equation}
which is the second term in the exponent of Eq.~\eqref{eq:after-average}.

Now we focus on $\mathbb{E}_M[\exp(M\text{-dependent exponent})]$. Reorganize the sum in the $M$-independent exponent by swapping the dummy indices $(a,b)$ in the second term gives
$
\sum_{a,b} M_{ba} \tilde X_a^\dag X_b = \sum_{a,b} M_{ab} \tilde X_b^\dag X_a,
$
and substituting this back gives
\begin{equation}
M\text{-dependent exponent} = \sum_{a,b} \left( M_{ab} X_a^\dag \tilde X_b - M_{ab} \tilde X_b^\dag X_a \right) = \sum_{a,b} M_{ab} \underbrace{ \left( X_a^\dag \tilde X_b - \tilde X_b^\dag X_a \right) }_{\coloneqq \  Q_{ab}}.
\end{equation}
So we must compute
\begin{equation}
\mathbb{E}_M \left[ \exp\left( \sum_{a,b} M_{ab} Q_{ab} \right) \right].
\end{equation}
Because all entries $M_{ab}$ are independently and identically distributed, the expectation of the product is the product of the expectations, giving
\begin{equation}\label{eq:expectation_M_exp}
\mathbb{E}_M \left[ \prod_{a,b} e^{M_{ab} Q_{ab}} \right] = \prod_{a,b} \mathbb{E}_{M_{ab}} \left[ e^{M_{ab} Q_{ab}} \right].
\end{equation}
Now, from the \emph{sparse} distribution from Eq.~\eqref{eq:sparse-ensemble} --- i.e.,  $\mathbb{P}(M) = (1-\frac{k}{N})\delta(M) + \frac{k}{N}p(M)$ --- we compute the single-site average $\mathbb{E}_{M_{ab}}[\cdot]$ for a single pair $(a,b)$:
\begin{align}\nonumber
\mathbb{E}_{M_{ab}} \left[ e^{M_{ab} Q_{ab}} \right] &= \int dM \, \mathbb{P}(M)  e^{M Q_{ab}} 
= \int dM \left[ \left(1 - \frac{k}{N}\right)\delta(M) + \frac{k}{N} \,p(M) \right] e^{M Q_{ab}} \\
&= \left(1 - \frac{k}{N}\right) \underbrace{\int dM  \delta(M) e^{M Q_{ab}}}_{=1} + \frac{k}{N} \int dM  p(M) e^{M Q_{ab}}.
\end{align}
The second integral is the definition of the characteristic function of the weight distribution $\hat p(t) = \int_{\mathbb{R}} p(w) e^{itw} dw$ evaluated at $t = -iQ_{ab}$:
\begin{equation}
\int dM  p(M) e^{M Q_{ab}} = \int dM  p(M) e^{i(-iQ_{ab})M} = \hat p(-i Q_{ab})
\end{equation}
Substituting these back, we get
\begin{align}
\mathbb{E}_{M_{ab}} \left[ e^{M_{ab} Q_{ab}} \right] = \left(1 - \frac{k}{N}\right) + \frac{k}{N} \hat p(-i Q_{ab})
=
1 + \frac{k}{N} \left( \hat p(-i Q_{ab}) - 1 \right).
\end{align}
Now we take the full product over all $N^2$ pairs $(a,b)$, so~\eqref{eq:expectation_M_exp} becomes
\begin{align}\nonumber
\prod_{a,b=1}^N \mathbb{E}_{M_{ab}} \left[ e^{M_{ab} Q_{ab}} \right] &= \prod_{a,b=1}^N \left[ 1 + \frac{k}{N} \left( \hat p(-i Q_{ab}) - 1 \right) \right]
=\exp\left( \sum_{a,b=1}^{N} \log\left[ 1 + \frac{k\left( \hat p(-i Q_{ab}) - 1 \right)}{N} \right] \right) \\
&\approx \exp\left( \sum_{a,b=1}^{N} \frac{k}{N} \left( \hat p(-i Q_{ab}) - 1 \right) \right),
\end{align}
where in the last step we use $\log(1+x) \approx x$ for small $x = \mathcal{O}(1/N)$, which is the case of the (sparse) finite connectivity $k$ such that $\frac{k}{N} = \mathcal{O}(1/N)$. This expression is exactly the integrand in Eq.~\eqref{eq:after-average} that involves the \emph{interaction kernel} of Eq.~\eqref{eq:C-kernel}, which is defined by simply equating
\begin{equation}
\exp\left[ -\frac{1}{N} \sum_{a,b=1}^N C(X_a, \tilde X_b) \right] = \exp\left[ \sum_{a,b=1}^N \frac{k}{N} \left( \hat p(-i Q_{ab}) - 1 \right) \right].
\end{equation}

\subsection{Superrotation invariance in the saddle point equations}
\label{app:proof-rotational-invariant-saddle}

We start with equation~\eqref{eq:pf-saddle-no-den} that we reproduce here for convenience:
\begin{equation}
  \label{sddlapp1}
g(X)= i  \int dX' d\tilde X \, C(X,\tilde X)\, e^{i\tilde g(\tilde X)+ig(X') - z X'^\dag \tilde X + \bar z\tilde X^\dag X'}.
\end{equation}
keeping in mind that
\begin{equation}
  C(X,X')
  = k\left[1-\int_{\mathbb{R}}p(w)\,e^{w(X^\dag X'-X'^\dag X)}dw\right].
  \label{eq:pf-C}
\end{equation}
We have assumed the \emph{superrotationally invariant} ansatz
\beq\label{gXX}
g(X)=g(X^\dag X).
\eeq

\paragraph*{Step 1: Introduce polar coordinates.}
Write bosonic variables in polar form as
\[
x=\sqrt{r}e^{i\alpha},\quad x'=\sqrt{\frac{z}{\bar z}r'\,}\,e^{i\alpha'+i\tilde\alpha},\quad
\tilde x=-i\sqrt{\tilde r\,}\,e^{i\tilde\alpha},\qquad
dX'=\frac{dr' d\alpha' d\bar\xi' d\xi'}{2\pi},
\]
and similarly for $d\tilde X$.
(Note that $r$ used here is the square of the usual radial variable.) We will set $\alpha=0$ without loss of generality, since $g$ is rotationally invariant and does not depend on $\alpha$.

For a superrotationally invariant $g$,
\beq\label{gTayl}
g(X^\dag X)=g(r+\bar\xi\xi)=g(r)+\del_r g(r)\bar\xi\xi,
\eeq
and similarly for $\tilde g$ and $g$ of other arguments.
Using \eqref{eq:pf-C}, the saddle-point equation \eqref{eq:pf-saddle-no-den} becomes
\begin{align}
  g(r)
  &= ik -ik\int_{\mathbb{R}}dw \,p(w)
     \int_0^\infty dr'\int_0^\infty d\tilde r
     \int_0^{2\pi}\frac{d\alpha'}{2\pi}\int_0^{2\pi}\frac{d\tilde\alpha}{2\pi}\,
     \mathcal{G}(r',\tilde r)\,
     \mathcal{B}(r,r',\tilde r;\alpha',\tilde\alpha), \label{eq:pf-before-Gr}
\end{align}
with the Grassmann integral factor
\begin{equation}
  \mathcal{G}
 \equiv\int d\bar\xi d\xi d\bar{\tilde\xi} d\tilde\xi
  \exp\Big[i \del_{\tilde r}\tilde g(\tilde r)\,\bar{\tilde\xi}\bar\xi+i\del_{r'}g(r')\,\bar\xi'\xi' - z \bar\xi'\tilde\xi + \bar z\bar{\tilde\xi} \xi'\Big],
  \label{eq:pf-GG}
\end{equation}
and the bosonic angular factor
\begin{equation}
  \mathcal{B}\equiv\exp\Big[i\tilde g(\tilde r)+ig(r') +2i |z|\sqrt{r'\tilde r}\cos\alpha'
-2iw\sqrt{r\tilde r}\cos\tilde\alpha
 \Big].
  \label{eq:pf-bos-ang}
\end{equation}
(On the left-hand side, we have extracted from $g(X^\dag X)$ given by (\ref{gTayl}) the bosonic part independent of the Grassmannian variables. Since $g$ only depends on one composite variable $r+\bar\xi\xi$, the purely bosonic part is enough to determine the function completely. Correspondingly, on the right-hand side, we have discarded terms coming from $C$ that depend on $\xi$ and $\bar\xi$.) 

\paragraph*{Step 2: Evaluating the Grassmannian and angular integrals.}

For evaluating the fermionic integral, we must expand the exponential in (\ref{eq:pf-GG}) in Taylor series and extract the terms that contain exactly one copy of each of $\bar\xi$, $\xi$, $\bar{\tilde\xi}$, $\tilde\xi$. These terms reside in the quadratic order of the Taylor expansion and amount to
\beq
\exp\Big[ \del_{\tilde r}\tilde g(\tilde r)\,\bar{\tilde\xi}\bar\xi-\del_{r'}g(r')\,\bar\xi'\xi' - z \bar\xi'\tilde\xi + \bar z\bar{\tilde\xi} \xi'\Big]
\to -\del_{\tilde r}\tilde g(\tilde r)\del_{r'}g(r')\,\bar{\tilde\xi}\bar\xi\bar\xi'\xi' -  \bar z z\,\bar\xi'\tilde\xi \bar{\tilde\xi} \xi'.
\eeq
To evaluate the fermionic integral, we must bring the $\xi$-variables to the canonical order $\tilde\xi\bar{\tilde\xi}\xi'\bar\xi'$, and the integral of this product is simply 1.
Implementing these operations, we obtain:
\label{lem:pf-Grassmann}
\begin{equation}
  \mathcal{G}(r',\tilde r') = \bar z z-\del_{\tilde r}\tilde g(\tilde r)\del_{r'}g(r').
  \label{eq:pf-Gr-closed}
\end{equation}

To evaluate the integrals over $\alpha'$ and $\tilde\alpha$ we employ the relation
\beq\label{J0id}
  \int_0^{2\pi}\frac{d\alpha}{2\pi}\,e^{iy\cos\alpha}=J_0\left(y\right),
\eeq
obtaining
\begin{equation}
  g(r)
  = ik -ik\int_{\mathbb{R}}dw \, p(w)\int_0^\infty \!\!dr'\int_0^\infty \!\!d\tilde r\,
      \Big[\bar z z-\del_{\tilde r}\tilde g(\tilde r)\del_{r'}g(r')\Big]
      e^{i\tilde g(\tilde r)+ig(r')}J_0(2|z|\sqrt{r'\tilde r}\,)\,
      J_0(2w\sqrt{ r\tilde r}).
  \label{eq:pf-after-angles}
\end{equation}

\paragraph*{Step 3: Integration by parts and cancellation of constants.}
We now integrate by parts in $r'$ and $\tilde r$ to trade $\del_{r'}\bar g(r')$ and $\del_{\tilde r}g(\tilde r)$ for
derivatives acting on the Bessel factors. Using
\begin{equation}
  \partial_{r'}J_0\left(2 w \sqrt{rr'}\right)
  = -\frac{w\sqrt{r}}{\sqrt{r'}}J_1\left(2w\sqrt{rr'}\right),
  \qquad
  \partial_{\tilde r}J_0\left(2|z|\sqrt{r'\tilde r}\right)
  = -\frac{|z|\sqrt{r'}}{\sqrt{\tilde r}}J_1\left(2|z|\sqrt{r'\tilde r}\right),
  \label{eq:pf-J-derivs}
\end{equation}
and integrating the second term in \eqref{eq:pf-after-angles} by parts first in $r'$ and then in
$\tilde r$, the boundary terms vanish:
\begin{align}
  g(r)
  &= ik -ik\int_{\mathbb{R}}dw\,  p(w)\int_0^\infty dr'\int_0^\infty d\tilde r\,
      |z|^2\, e^{i\tilde g(\tilde r)+ig(r')}
      J_0\left(2w\sqrt{rr'}\right)
      J_0\left(2|z|\sqrt{r'\tilde r}\right)
      \nonumber\\[-2pt]
  &\quad
      - ik\int_{\mathbb{R}}dw \, p(w)\int_0^\infty dr'\int_0^\infty d\tilde r\,
   e^{i\tilde g(\tilde r)+ig(r')}\del_{r'}\partial_{\tilde r}
      \Big[ J_0\left(2|w|\sqrt{rr'}\right)
            J_0\left(2|z|\sqrt{r'\tilde r}\right)\Big].
  \label{eq:pf-after-IBP-raw}
\end{align}
Next, we use the Bessel differential equation identity,
$(x^2\partial_x^2 + x\partial_x + x^2)J_0(x)=0$, with $x=2|z|\sqrt{r'\tilde r}$ to rewrite
the $\bar z z$ term as a total derivative in $(r',\tilde r)$, acting with
$\partial_{r'}$ and $\partial_{\tilde r}$ using \eqref{eq:pf-J-derivs} and chain rule.
Explicitly,
\begin{equation}
  \bar z zJ_0\left(2|z|\sqrt{r'\tilde r}\right)
  = \frac{\sqrt{\tilde r}}{4\sqrt{r'}}\partial_{r'}\Big[\sqrt{r'\tilde r}
         \partial_{\tilde r} J_0\left(2|z|\sqrt{r'\tilde r}\right)\Big]
       + \frac{\sqrt{r'}}{4\sqrt{\tilde r}}\partial_{\tilde r}\Big[\sqrt{r'\tilde r}
         \partial_{r'} J_0\left(2|z|\sqrt{r'\tilde r}\right)\Big].
  \label{eq:pf-bessel-total-der}
\end{equation}
Inserting \eqref{eq:pf-bessel-total-der} into the first line of \eqref{eq:pf-after-IBP-raw} and
integrating by parts, the entire $\bar z z$ contribution cancels against the boundary pieces generated
by the second line. Then, using \eqref{eq:pf-J-derivs} again yields
\begin{equation}
  g(\rho)
  =
 k r\int_{\mathbb{R}}dw\,w\,p(w)\int_0^\infty dr'\int_0^\infty d\tilde r\,\,
  \frac{J_1\left(2w\sqrt{rr'}\right)}{\sqrt{rr'}}\,
  J_0\left(2|z|\sqrt{r'\tilde r}\right)
  e^{i\tilde g(\tilde r)+ig(r')}\,\partial_{\tilde r}g(\tilde r).
  \label{eq:pf-final-saddle}
\end{equation}

\subsection{Superrotation invariance and the resolvent}\label{app:proof-rotational-invariant-G}

We need to simplify the expression (\ref{Gg}),
\beq
G(z)=
 \displaystyle \int dX d\tilde X\,\bar x\tilde x \,e^{i\tilde g(\tilde X)+ig(X) - z X^\dag \tilde X + \bar z\tilde X^\dag X}
\eeq
under the assumption $g(X)=g(X^\dag X)$. 
We start with the fermionic integrals by writing (throughout, $g'$ will denote the derivative of $g(X^\dag X)$ with respect to its $X^\dag X$ argument). First,
\beq
e^{i\tilde g(\tilde X^\dag \tilde X)+ig(X^\dag X) - z X^\dag \tilde X + \bar z\tilde X^\dag X}=e^{i\tilde g(\bar{\tilde x}\tilde x)+ig(\bar x x) - z \bar x \tilde x + \bar z\bar{\tilde x} x}\,
e^{i\tilde g'(\bar{\tilde x}\tilde x)\,\bar{\tilde \xi}\tilde\xi+ig'(\bar xx)\,\bar\xi\xi - z \bar\xi \tilde \xi + \bar z\bar{\tilde \xi} \xi}.
\eeq
For evaluating the fermionic integral, we must expand the second exponential in Taylor series and extract the terms that contain exactly one copy of each of $\bar\xi$, $\xi$, $\bar{\tilde\xi}$, $\tilde\xi$. These terms reside in the quadratic order of the Taylor expansion and amount to
\beq
e^{i\tilde g(\tilde X^\dag \tilde X)+ig(X^\dag X) - z X^\dag \tilde X + \bar z\tilde X^\dag X}\to e^{i\tilde g(\bar{\tilde x}\tilde x)+ig(\bar x x) - z \bar x \tilde x + \bar z\bar{\tilde x} x}
\left[-\tilde g'(\bar{\tilde x}\tilde x)g'(\bar x x)\,\bar\xi\xi\bar{\tilde \xi}\tilde\xi-|z|^2\bar\xi \tilde \xi \bar{\tilde \xi} \xi\right].
\eeq
To evaluate the fermionic integral, we must bring the $\xi$-variables to the canonical order $\tilde\xi\bar{\tilde\xi}\xi\bar\xi$, and the integral of this product is simply 1.
Implementing these operations, we obtain:
\beq
G(z)= \frac1{\pi^2}\int d\bar x dx d\bar{\tilde x} d\tilde x\, \bar x\tilde x\, e^{i\tilde g(\bar{\tilde x}\tilde x)+ig(\bar x x) - z \bar x \tilde x + \bar z\bar{\tilde x} x}
\left[|z|^2-g'(\bar x x)g'(\bar{\tilde x}\tilde x)\right].
\eeq
The remaining bosonic integrals are best rewritten in polar coordinates. Since the integrand is invariant under simultaneous rotations of $x$ and $\tilde x$ by a complex phase $e^{i\alpha_0}$, we can use this symmetry to align $x$ with the real axis and simultaneously multiply the integral by $2\pi$ due to the trivial integration over $\alpha_0$. We thus write:
\beq
x=\sqrt{r},\quad \tilde x=\sqrt{\tilde r}\,e^{i\alpha},\qquad  d\bar x dx d\bar{\tilde x} d\tilde x\to \frac14\,2\pi\, drd\tilde r d\alpha,
\eeq
and hence
\beq
G(z)= \frac1{2\pi}\int drd\tilde r d\alpha\, \sqrt{r\tilde r} e^{i\alpha}\,\Big[|z|^2-g'(r)\tilde g'(\tilde r)\Big] \,\exp\left[i\tilde g(\tilde r)+ig(r) - z \sqrt{r\tilde r} e^{i\alpha} + \bar z\sqrt{r\tilde r} e^{-i\alpha}\right].
\eeq
It is convenient to write $iz=|z|e^{i\phi}$ and introduce a new variable $\alpha$ equal $\alpha+\phi$ in the above expression. Then,
\beq
G(z)= \frac{e^{-i\phi}}{2\pi}\int drd\tilde r d\alpha\, \sqrt{r\tilde r} e^{i\alpha}\,\Big[|z|^2-g'(r)\tilde g'(\tilde r)\Big] \,\exp\left[i\tilde g(\tilde r)+ig(r) + 2i|z| \sqrt{r\tilde r} \cos\alpha\right].
\eeq
By taking one half of the sum of this integral and the same integral with $\alpha\to-\alpha$, $e^{i\alpha}$ gets replaced with $\cos\alpha$, and then we can employ the following representation of the Bessel function $J_1(y)=(2\pi i)^{-1}\int_0^{2\pi} d\alpha\, \cos\alpha\, e^{iy\cos\alpha}$, obtaining
\beq
G(z)=\sqrt{\frac{\bar z}z}\int drd\tilde r  \,\sqrt{r\tilde r} \,J_1(2|z| \sqrt{r\tilde r})\,\Big[|z|^2-\del_rg(r)\del_{\tilde r}\tilde g(\tilde r)\Big] \,e^{i\tilde g(\tilde r)+ig(r)}.
\eeq


\section{Dense two-population limit as a special case}
\label{app:RAM-like_model}

We show how the field-theoretic machinery developed in the main text reproduces the classical
Rajan-Abbott (RA) two-population spectrum (where the zero-row-sum balance condition is imposed \emph{on average} rather than as a strict constraint) \cite{Rajan2006} as a special case. We work in the dense limit. Within our framework, we set homogeneous timescales and center the spectrum by removing the trivial $-I$ shift, and take 
\[
\tau_i\equiv 1,\qquad \tilde J :=J+I=W\mathcal{H},
\]
so that the bulk of $\tilde J $ is centered at the origin and $\mathrm{Spec}(J)=\mathrm{Spec}(\tilde J )-1$.
Each presynaptic neuron (column) is either inhibitory or excitatory with probabilities
$f$ and $1-f$, respectively. Conditional on the column type, the \emph{random} weights are i.i.d. with
\[
W_{ij}\sim \mathcal{N}\left(\frac{\mu_T}{N},\frac{\sigma^2}{N}\right),
\qquad T\in\{I,E\},
\]
and the column gain is
\[
h_j =
\begin{cases}
1, & j\in I,\\[2pt]
\alpha^{-1/2}, & j\in E,
\end{cases}
\qquad\Longrightarrow\qquad
\mathbb{E}_h[h^2]=f\cdot 1+(1-f)\cdot \alpha^{-1}.
\]
The row-sum (mean) constraint is imposed only \emph{on average},
$
f\mu_I+(1-f)\mu_E=0
$,
so the bulk is unchanged but a rank-one outlier may appear. Namely,
if the zero-row-sum constraint is imposed only on the mean rather than exactly
row by row, a rank-one component induces an $\mathcal{O}(1)$ outlier eigenvalue whose weight in the empirical
density is $1/N$. This outlier does not affect the bulk disk in the limit  $N\to\infty$ and is therefore
negligible for the mean spectral density (see, for example, \cite{Rajan2006}).

\paragraph*{Dense-limit saddle with Gaussian disorder.}
With Gaussian disorder of variance $\sigma^2/N$ and the superrotation-invariant ansatz
$g(X)=aX^\dag X$ (Sec.~\ref{sec:intro_susy}), the saddle point equation reduces to
\beq
a  = -\int dh p_h(h) \frac{a}{\tilde{a} a - \bar z z/h^2} = -\Bigg[\frac{(1-f)a/\alpha}{\tilde{a} a/\alpha - \bar z z} +
\frac{fa}{\tilde{a} a - \bar z z}\Bigg],\label{eq:RA_saddle_a}
\eeq
where $z$ is the spectral parameter for the centered matrix $\tilde J =W\mathcal H$.
The trivial solution $a=0$ holds outside the spectrum. The non-trivial solution is
determined by \eqref{eq:RA_saddle_a} with identification $\tilde{a}=-\bar{a}$ as in \eqref{utildeubar}. This is the two-population specialization of the general dense saddle discussed in the main text, with $h\in\{1,\alpha^{-1/2}\}$.

\begin{figure}[t]
    \centering
    \includegraphics[width=0.8\linewidth]{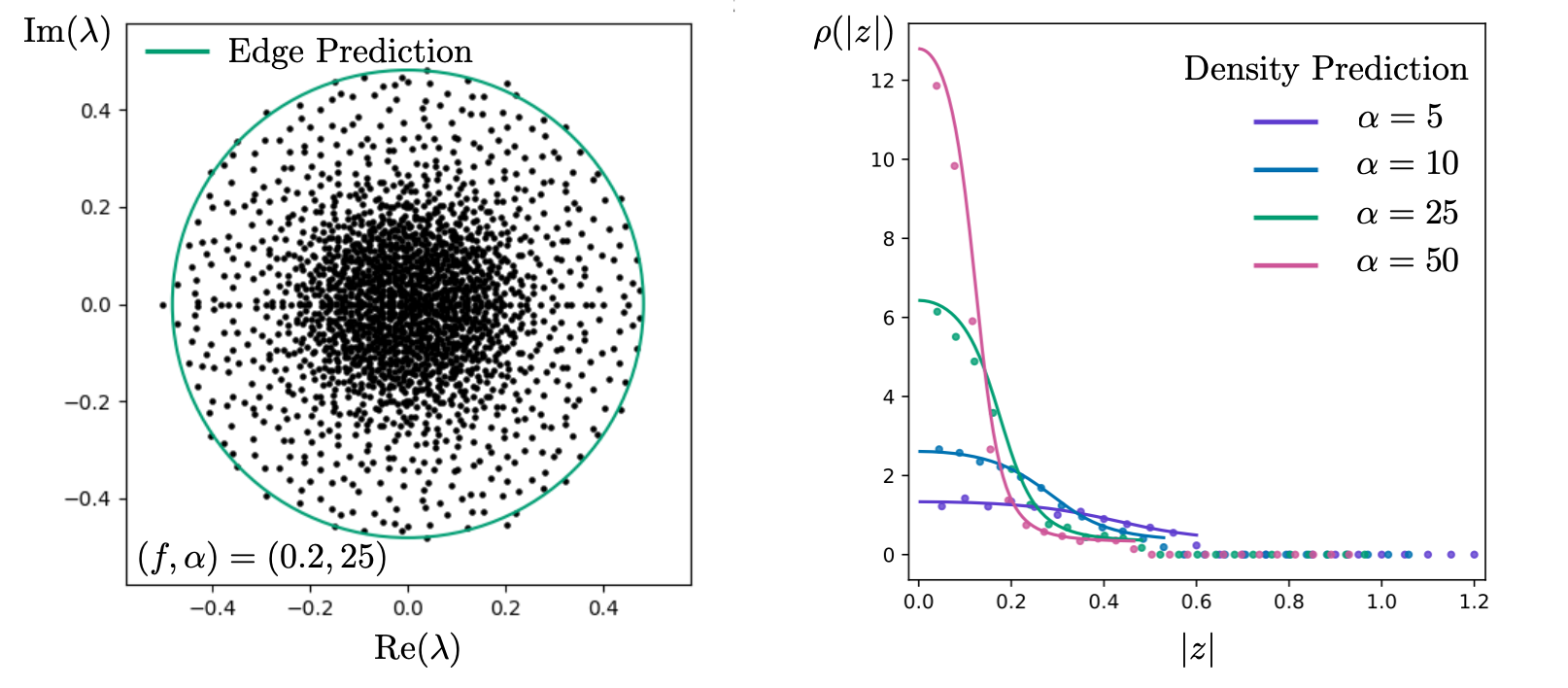}
    \caption{
    \textbf{Validation of the dense two-population (Rajan-Abbott) spectral distribution against numerical spectra} from exact diagonalization with $N=2000$, $\sigma=1$, and an inhibitory fraction $f=0.2$. 
    The mean weights are balanced to satisfy $f\mu_I + (1-f)\mu_E = 0$ on average, using $\mu_I = -1$ and $\mu_E = 0.25$.
    \textbf{(Left)} Numerical eigenvalue spectrum (black dots) for a single matrix realization with $\alpha=25$. The solid teal line is the theoretical spectral edge predicted by Eq.~\eqref{eq:RA_radius}, $R=\sigma\sqrt{f+(1-f)/\alpha}$.
    \textbf{(Right)} Radial spectral density $\rho(|z|)$. Dots are numerical density estimated from multiple samples. Solid lines show the non-uniform theoretical density predicted by Eq.~\eqref{eq:RA_density} for various $\alpha \in \{5, 10, 25, 50\}$, each terminating at its edge, with respective theoretical radius $R$.
    }
    \label{fig:RAM_spectrum}
\end{figure}
\paragraph*{Spectral edge and disk radius.}
From \eqref{eq:RA_saddle_a}, there is a trivial solution $a=0$, and a nontrivial solution of the form
\begin{align}
    \bar{a}a=\frac{1-\bar{z}z(1+\alpha)\pm\sqrt{\left[(1-\alpha)\bar{z}z+1\right]^{2}-4f(1-\alpha)\bar{z}z}}{2}.
\end{align} 
The spectral edge can then be determined by the point where these two solutions are matched. This can be solved by the set of $(z, \bar{z})$ satisfying \eqref{eq:RA_saddle_a}, such that
\beq
    1=\left(\frac{(1-f)/\alpha}{\bar{a}a/\alpha+\bar{z}z}+\frac{f}{\bar{a}a+\bar{z}z}\right)\lvert_{\bar{a},a=0},\qquad
    \bar{z}z=f+(1-f)/\alpha.
\eeq
Thus the bulk of $\tilde J $ is the disk
\begin{equation}
\label{eq:RA_radius}
\mathrm{Spec}_{\mathrm{bulk}}(\tilde J ) = \{z\in\C: |z|\le R\},\qquad
R = \sigma\sqrt{f+(1-f)/\alpha}  = \sigma\sqrt{\mathbb{E}_h[h^2]}. 
\end{equation}
Equivalently, the spectrum of the original Jacobian $J=-I+W\mathcal H$ is the same disk shifted by $-1$
along the real axis. Eq.~\eqref{eq:RA_radius} coincides with the RA radius \cite{Rajan2006}.

\paragraph*{Bulk resolvent and density.}
For Gaussian disorder the SUSY action is quadratic and the interior saddle takes the linear form
$g(X)=aX^\dag X$. Denote $\omega:= \bar z z$.
Evaluating the single-site ratio at this saddle gives the interior resolvent in the form
\begin{equation}
\label{eq:RA_resolvent_phi}
G(z)=\frac{1}{N}\mathbb{E}\Tr(\tilde J -zI)^{-1}
\sim \int \frac{dh}{h^2} p_h(h) \frac{i \bar z}{\bar a a + \bar z z /h^2} = \bar z\phi(\omega),
\qquad
\phi(\omega)=\frac{f}{\bar a a+\omega} + \frac{(1-f)}{\bar a a/\alpha+\omega},
\end{equation}
where $a(\omega)$ is determined implicitly by \eqref{eq:RA_saddle_a}. The spectral density then follows from using $\partial_{\bar z}(\bar z\phi(\omega))=\phi(\omega)+\omega\phi'(\omega)$:
\begin{equation}
\label{eq:RA_density}
\rho_(z)=\frac{1}{\pi}\partial_{\bar z}G(z)
=\frac{1}{\pi}\Big[\phi(\omega)+\omega\phi'(\omega)\Big],
\qquad \omega=|z|^2,
\end{equation}
supported on the disk $|z|\le R$ in \eqref{eq:RA_radius}. (Note again that at the edge $\bar a a \to 0$ and $G$ matches the exterior
holomorphic branch, as in the circular law.) The analytic derivations are compared with the numerical spectra in Fig.~\ref{fig:RAM_spectrum}.

\paragraph*{Relation to the main-text edge formula.}
The dense-limit edge condition derived in the main text reads
$
\sigma^2\mathbb{E}_h[h^2]\mathbb{E}_\tau[|z\tau+1|^{-2}]=1
$.
For the centered matrix $\tilde J $ with $\tau\equiv 1$ (and no $-I$ shift) this reduces to
$
\sigma^2\mathbb{E}_h[h^2]/|z|^2=1
$,
which immediately gives \eqref{eq:RA_radius}.

\end{document}